\documentclass[%
superscriptaddress,
twocolumn, 
longbibliography,
 amsmath,amssymb,
 aps,
pra,
floatfix,
]{revtex4-2}

\usepackage{graphicx}
\usepackage{dcolumn}
\usepackage{bm}
\usepackage[mathlines]{lineno}
\usepackage{seqsplit} 
\usepackage{physics}
\usepackage{bbm}
\usepackage{color}
\usepackage{comment}
\usepackage{amsthm}
\usepackage{amssymb}
\usepackage{grffile}
\theoremstyle{plain}
\newtheorem{theorem}{Theorem}
\theoremstyle{remark}

\theoremstyle{plain}

\theoremstyle{definition}
\newtheorem{definition}{Definition}
\theoremstyle{remark}

\usepackage{seqsplit}

\newcommand{\subfigimg}[3][,]{%
	\setbox1=\hbox{\includegraphics[#1]{#3}}
	\leavevmode\rlap{\usebox1}
	\rlap{\hspace*{2pt}\raisebox{\dimexpr\ht1-0.5\baselineskip}{{\bfseries \large\textsf{#2}}}}
	\phantom{\usebox1}
}

\usepackage[colorlinks=true, urlcolor=blue, citecolor=blue,linkcolor=blue,citebordercolor={1 0 0},linkbordercolor={0 0 1}]{hyperref}
\usepackage{listings}
\usepackage{parskip} 
\usepackage{graphicx}
\graphicspath{{./Figures/}}

\renewcommand{\eqref}[1]{Eq.~(\ref{#1})} 
\newcommand{\figref}[1]{Fig.~\ref{#1}} 

\usepackage{booktabs}

\usepackage{csquotes}
\MakeOuterQuote{"}
\renewcommand{\vec}[1]{\bm{#1}}

\definecolor{KB}{rgb}{0.4,0.3,0.9}

\definecolor{THc}{rgb}{0.9,0.3,0.2}

\definecolor{Change}{rgb}{0.1,0.1,0.8}

\definecolor{KH}{rgb}{0.2,0.3,0.9}

\usepackage{float}

\begin{document}


\title{Fast-Forwarding with NISQ Processors without Feedback Loop}

\author{Kian Hwee Lim}	
\email{kianhwee\_lim@u.nus.edu}
\affiliation{Centre for Quantum Technologies, National University of Singapore 117543, Singapore}

\author{Tobias Haug}
\affiliation{QOLS, Blackett Laboratory, Imperial College London SW7 2AZ, UK}

\author{Leong Chuan Kwek}
\affiliation{Centre for Quantum Technologies, National University of Singapore 117543, Singapore}
\affiliation{National Institute of Education, Nanyang Technological University, 1 Nanyang Walk, Singapore 637616}
\affiliation{MajuLab, CNRS-UNS-NUS-NTU International Joint Research Unit, UMI 3654, Singapore}

\author{Kishor Bharti}
\email{kishor.bharti1@gmail.com}
\affiliation{Centre for Quantum Technologies, National University of Singapore 117543, Singapore}



\begin{abstract}
  Simulating quantum dynamics is expected to be performed more easily on a
  quantum computer than on a classical computer.  However, the currently
  available quantum devices lack the capability to implement fault-tolerant
  quantum algorithms for quantum simulation. Hybrid classical quantum algorithms
  such as the variational quantum algorithms have been proposed to effectively
  use current term quantum devices.  One promising approach to quantum
  simulation in the noisy intermediate-scale quantum (NISQ) era is the
  diagonalisation based approach, with some of the promising examples being the
  subspace Variational Quantum Simulator (SVQS), Variational Fast Forwarding
  (VFF), fixed-state Variational Fast Forwarding (fs-VFF), and the Variational
  Hamiltonian Diagonalisation (VHD) algorithms. However, these algorithms
  require a feedback loop between the classical and quantum computers, which can
  be a crucial bottleneck in practical application.  Here, we present the
  Classical Quantum Fast Forwarding (CQFF) as an alternative diagonalisation
  based algorithm for quantum simulation.  CQFF shares some similarities with
  SVQS, VFF, fs-VFF and VHD but removes the need for a classical-quantum
  feedback loop and controlled multi-qubit unitaries.  The CQFF algorithm does
  not suffer from the barren plateau problem and the accuracy can be
  systematically increased. Furthermore, if the Hamiltonian to be simulated is
  expressed as a linear combination of tensored-Pauli matrices, the CQFF
  algorithm reduces to the task of sampling some many-body quantum state in a
  set of Pauli-rotated bases, which is easy to do in the NISQ era.  We run the
  CQFF algorithm on existing quantum processors and demonstrate the promise of
  the CQFF algorithm for current-term quantum hardware. We compare CQFF with
  Trotterization for a $XY$ spin chain model Hamiltonian and find that the CQFF
  algorithm can simulate the dynamics more than $10^5$ times longer than
  Trotterization on current-term quantum hardware. This provides a  $10^4$ times
  improvement over the previous record.
\end{abstract}

\maketitle

\section{Introduction}
In the 1980s Feynman suggested that since nature is quantum-mechanical, it
would be easier to simulate a quantum system on a quantum computer rather
than a classical computer~\cite{feynman1982simulating}.  His idea has
far-reaching implications on fields such as chemistry and materials science.
Many quantum simulation algorithms have been proposed since then, with
Trotterization being one of the most prominent \cite{lloyd1996universal}.
Trotterization, however, requires an extensive amount of quantum resources
and most likely can only be implemented on fault-tolerant quantum
computers~\cite{poulin2014trotter}.  Hence, in the current noisy
intermediate-scale quantum (NISQ)~\cite{Preskill2018quantumcomputingin} era
when we do not yet have fault-tolerant quantum computers, alternate methods
have to be used.  In the NISQ era, most algorithms make use of a classical
quantum feedback loop \cite{bharti_noisy_2021,cerezo_variational_2020}.  In
each iteration, a classical computer runs a classical optimization routine to
determine a set of parameters for a parameterized quantum circuit to minimize
an appropriate cost function.  On the quantum computer, the cost function is
calculated and then sent as input to the classical computer for the next
iteration.  The canonical example of such a quantum algorithm is the
variational quantum eigensolver~\cite{peruzzo2014variational,
kandala2017hardware}.  For the task of quantum simulation, various algorithms
have been proposed, such as the Variational Quantum Simulator (VQS)
\cite{li_efficient_2017,yuan_theory_2019}, Subspace Variational Quantum
Simulator (SVQS) \cite{heya_subspace_2019}, Quantum Assisted Simulator (QAS)
\cite{bharti2020simulator,lau2021quantum}, Variational Fast Forwarding (VFF)
\cite{cirstoiu_variational_2020}, fixed-state Variational Fast Forwarding
(fs-VFF) \cite{gibbs_long-time_2021}, Generalised Quantum Assisted Simulator
(GQAS) \cite{haug_generalized_2020}, Variational Hamiltonian Diagonalisation
(VHD) \cite{commeau2020variational}, projected-Variational Quantum Dynamics
(p-VQD) \cite{barison2021efficient} and the truncated Taylor quantum
simulator (TTQS) \cite{lau2021nisq}.  These algorithms allow to simulate
quantum dynamics beyond the coherence time possible with Trotterization in
the NISQ era.

In this work, we focus on diagonalisation based approaches, i.e.  the SVQS,
VFF, fs-VFF and VHD algorithms.  The "no fast-forwarding theorem" tells us
that for a quantum system, simulating the time evolution with respect to a
generic Hamiltonian $H$ for time $T$ requires at least a number of gates that
scales linearly with $T$, which means that in general it is not possible to
perform quantum simulation with a sublinear amount of resources
\cite{amChildsLimitations,berry2007efficient}.  It has been shown that a
given Hamiltonian can be fast-forwarded if and only if it corresponds to
violations of the time-energy uncertainty relations and equivalently allows
for precise energy measurements \cite{atia2017fast}.  For a discussion on the
implications of asymptotic fast-forwarding on quantum simulation with NISQ
devices, refer to \cite{cirstoiu_variational_2020}.

The main idea of the NISQ quantum simulation algorithms based on
diagonalisation is to variationally find a unitary transformation into a
space such that the time evolution can be easily performed with a fixed
circuit structure.  We now proceed to review the aforementioned algorithms.
First, for the SVQS algorithm, the idea is to first variationally search for
a unitary transformation into the subspace spanned by the low-lying energy
eigenstates of the Hamiltonian $H$.  Then, the time evolution in that
subspace can be easily done by just single-qubit $Z$-rotations on each qubit.
Next, for the VFF algorithm, the idea is to first variationally find the
unitary transformation to diagonalise the time evolution operator
$e^{-iH\Delta t}$ with a small timestep $\Delta t$.  Then, one applies the
diagonal evolution operator multiple times, which can be done without
requiring additional resources, allowing one to fast-forward to large
evolution times.  The fs-VFF algorithm is a modification of the VFF
algorithm, where the observation is made that if the initial state to be
evolved $\ket{\psi_0}$ lies in the span of $n_{eig}$ energy eigenstates, then
$e^{-iHT} \ket{\psi_0}$ lies in the same span.  In the fs-VFF algorithm,
diagonalisation of $e^{-i H \Delta t}$ is not done over the entire Hilbert
space, but only on the $n_\text{eig}$ dimensional subspace that
$\ket{\psi_0}$ lies in.  Finally, for the VHD algorithm, the idea is to first
variationally find the unitary transformation to diagonalise the Hamiltonian
operator $H$.  Then, $H$ can be easily exponentiated to obtain $e^{-iHT}$,
which which can be done without requiring additional resources.  For these
diagonalisation based variational quantum simulation algorithms, the bulk of
the work is to approximately variationally diagonalise an operator.

The variational approaches described above suffer from a few weaknesses, namely:
\begin{enumerate}
  \item{For variational quantum algorithms,
  the classical quantum feedback loop can be a major bottleneck when running
  the algorithm on current cloud-based quantum computers, as each job for the
  quantum computer has to wait in a queue. For each iteration, one needs to
  wait for the result from the quantum computer, which can take an extensive
  amount of time. The SVQS algorithm requires two classical quantum feedback
  loops, one for the initial state preparation and another for variationally
  searching for the unitary transform into the space spanned by the low lying
  energy eigenstates. The VFF/fs-VFF and the VHD algorithms require one
  classical quantum feedback loop.}
  \item{For variational quantum algorithms, there can potentially be barren
  plateaus when number of qubits, hardware noise or entanglement
  increases~\cite{mcclean2018barren,
  huang2019nearterm,sharma2020trainability,wang2021noiseinduced,
  cerezo2020costfunctiondependent,haug2021capacity}, which lead to an
  exponential decrease in the variance of the gradient and rendering training
  of these variational approaches very challenging.}
  \item{The fs-VFF algorithm requires the Hadamard test in the computation of
  $n_{eig}$, the VHD algorithm requires the Hadamard test in the computation
  of the cost function and the gradients, and the VFF algorithm requires the
  use of the Local Hilbert-Schmidt test.  The Hadamard test is hard to do in
  the NISQ era due to it requiring controlled multi-qubit unitaries.  The
  Local Hilbert-Schmidt test does not require controlled multi-qubit
  unitaries but it requires many controlled single-qubit unitary gates
  between qubits which are physically far separated.  Depending on the type
  of NISQ quantum computer, this operation may require many SWAP gates and
  could be a potential bottleneck.}
\end{enumerate}

Here, we propose another diagonalisation based algorithm which we call the
Classical Quantum Fast Forwarding (CQFF) algorithm.  Our algorithm solves all
of the aforementioned challenges.  Some of the features of our algorithm are:
\begin{enumerate}
  \item{The CQFF algorithm has no classical quantum feedback loop unlike the
  SVQS, VFF, fs-VFF and VHD algorithms.}
  \item{The CQFF algorithm avoids the barren plateau problem as there is no
  parametrized quantum circuit that is being updated.}
  \item{The CQFF algorithm has a systematic way of constructing the ansatz
  and the simulation result can always be improved by considering a higher
  value of $K$ when computing $\mathbb{C}\mathbb{S}_K$.}
  \item{For the CQFF algorithm, the quantum processor's output can be easily
  computed without requiring any controlled multi-qubit unitaries, such as
  those required in the Hadamard test.}
\end{enumerate}

\section{The CQFF algorithm}
Our ansatz is a hybrid state, which is a classical combination of $L$ quantum
states
\begin{equation}
  \label{eqn: hybrid ansatz equation}
  \ket{\psi(\boldsymbol{\alpha}(t))}=\sum_{i=1}^L\alpha_i(t)\ket{\chi_i}\,,
\end{equation}
where $\boldsymbol{\alpha}(t)\in\mathbb{C}^L$ and $\{\ket{\chi_i}\}_{i=1}^L$
is a set of $L$ quantum states.  We now want to evolve a state
$\ket{\psi(\boldsymbol{\alpha}(t=0))}$ for a time $T$ under a given
Hamiltonian $H$ with $i\partial_t\ket{\psi(t)}=H\ket{\psi(t)}$.  We assume
that the Hamiltonian $H$ that we want to simulate is given as a linear
combination of $r$ unitaries
\begin{equation}
  \label{eqn: Hamiltonian as linear combination of unitaries}
  H = \sum_{i=1}^r \beta_i U_i\,,
\end{equation}
where $\beta_i \in \mathbb{C}$ and the $N$-qubit unitaries $U_i
\in\text{SU}(2^N \equiv \mathcal{N})$, for $i \in \{1,2,\dots r\}$. Moreover,
each unitary acts non-trivially on at most $\mathcal{O}(poly(logN))$ qubits.
If the unitaries in \eqref{eqn: Hamiltonian as linear combination of
unitaries} are tensored-Pauli matrices, then we do not need the
$\mathcal{O}(poly(logN))$ constraint.

Now, we have to choose the states $\ket{\chi_i}$ in our hybrid ansatz.  Our
goal is that the states $\{\ket{\chi_i}\}$ span the space of the evolution.
We use a NISQ friendly approach first put forward
in~\cite{bharti2020iterative}, which is based on the idea of Krylov
expansion.  Given a matrix $A$, vector $b$ and some scalar $\tau$, $\exp(\tau
A) b$ can be approximated as some order $m$ polynomial which can be
reformulated as an element in the following Krylov
subspace~\cite{saad1992analysis},
\begin{equation}
{\cal{K}}_m = \text{span}\left\{ b, Ab, \cdots, A^{m-1}b\right\}.
\end{equation}

One can improve the approximation accuracy by increasing $m$.
For the case where $A$ is the Hamiltonian and $b$ is a quantum state, the
Krylov subspace based approximation for imaginary or real time evolution is a
natural choice.  This motivates us to construct the states $\ket{\chi_i}$ of
our hybrid ansatz using the $r$ unitaries $U_i$ in the Hamiltonian in
\eqref{eqn: Hamiltonian as linear combination of unitaries}.  We choose
states $\ket{\chi_i}$ from the set $\mathbb{C}\mathbb{S}_K$, which we call
the set of cumulative $K$-moment states.  Briefly speaking, we have
$\mathbb{C}\mathbb{S}_K = \mathbb{S}_0 \cup \mathbb{S}_1 \cup \dots \cup
\mathbb{S}_K$, where $\mathbb{S}_0 = \{\ket{\phi}\}$, and $\mathbb{S}_p$ for
$1 \leq p \leq K$ is defined as
\begin{equation}
  \mathbb{S}_p = \{U_{i_p} \dots U_{i_2} U_{i_1} \ket{\phi}\}_{i_1=1,\dots
  i_p=1}^r\,,
\end{equation}
where the unitaries $U_{i_a}$ are those in the Hamiltonian in \eqref{eqn:
Hamiltonian as linear combination of unitaries}.  We assume that the state
$\ket{\phi}$ can be efficiently prepared on a quantum computer.  A formal
definition and examples of $\mathbb{C}\mathbb{S}_K$ are given in
Appendix~\ref{Appendix: CSK definition}. We also justify
in Appendix~\ref{Appendix: CSK definition} how the ansatz states
$\ket{\chi_i}$ from the set $\mathbb{C}\mathbb{S}_K$ capture the time
evolution with a given Hamiltonian. The initial state for $t=0$ can be the
quantum state prepared on the quantum computer with $\vec{\alpha}(t=0)$ such
that $\ket{\psi(\vec{\alpha}(t=0)} = \ket{\phi}$, or more generally we can
choose $\vec{\alpha}(t=0)$ as linear combination of states in
$\mathbb{C}\mathbb{S}_K$.  Note that $\ket{\phi}$ is the only state that
needs to be prepared on the quantum computer.

Next, we compute the $D$ and $E$ matrices on the quantum computer, where the
matrix elements of $D$ and $E$ matrices  are given by
\begin{align}
  \label{eqn: D matrix element}
  D_{ij} &= \mel{\chi_i}{H}{\chi_j} = \sum_{a=1}^r \beta_a \mel{\chi_i}{U_a}{\chi_j} \\
  \label{eqn: E matrix element}
  E_{ij} &= \braket{\chi_i}{\chi_j}\,.
\end{align}
We note that if the unitaries $U_i$ are just tensored-Pauli matrices, then
the calculation of these matrix elements just reduces to the problem of
sampling the state $\ket{\phi}$ in some Pauli-rotated basis.  Otherwise,
since the unitaries $U_i$ acts trivially only on $\mathcal{O}(poly(log(N)))$
qubits, we can use the methods in~\cite{Mitarai} to compute the expectation
values without the need for Hadamard tests or complicated controlled
multi-qubit unitaries.  The only task of the quantum computer is to calculate
$D$ and $E$ matrix.  Thus, if the unitaries $U_i$ are just tensored-Pauli
matrices, then we have mapped the task of Hamiltonian simulation to a circuit
sampling task.

Once we have the $D$ and $E$ matrices, the job of the quantum computer is
done; what remains is the classical post-processing stage with three steps.
We give a short summary of the classical post-processing stage. Firstly, we
use the $D$ and $E$ matrices to compute a diagonal representation of $H$.
Next, we use the diagonal representation of $H$ to trivially compute a
diagonal representation of $e^{-i H t}$. Lastly, we perform the time
evolution to find the vector $\vec{\alpha}(t)$ in \eqref{eqn: hybrid ansatz
equation}.

We now detail the post-processing steps.  The evolution of the hybrid ansatz,
which we constructed using the evolution Hamiltonian $H$, is assumed to be
approximately constrained within the space spanned by
$\mathbb{C}\mathbb{S}_K$.  We define the projected Hamiltonian
$[H]_{\mathbb{C}\mathbb{S}_K}$ (see Appendix~\ref{Appendix: derivation of
diagonal representation} for full derivation) that projects the full $H$ onto
the space spanned by hybrid state ansatz given by the cumulative $K$-moment
states $\mathbb{C}\mathbb{S}_K$
\begin{equation}
  \label{eqn: H w.r.t CSK}
  [H]_{\mathbb{C}\mathbb{S}_K} = \sum_i \lambda_i \vec{v}_i \vec{v}_i^\dag E\,,
\end{equation}
where $\lambda_i$ is the $i$-th eigenvalue and $\vec{v}_i$ the $i$-th
eigenvector of the generalized eigenvalue problem defined below
\cite{bharti2020iterative,jia2001analysis,generalisedEigvalueSolverCitation1}
\begin{equation}
  \label{eqn: generalised eigenvalue problem}
  D\vec{v} = \lambda E \vec{v}.
\end{equation}
Note that in general $E$ may not be full rank, however the eigenvectors
corresponding to the nullspace of $E$ do not contribute in the subsequent
equations in the paper and can be safely ignored (see Appendix~\ref{Appendix:
derivation of diagonal representation}).  The above generalised eigenvalue
problem is related to the following Quadratically Constrained Quadratic
Program (QCQP)
\begin{gather}
  \min_{\vec{v}}(\vec{v}^\dag D \vec{v}) \nonumber \\ 
  \label{eqn: minimisation program}
  \text{subject to } \vec{v}^\dag E \vec{v} = 1,
\end{gather}
which is a well characterised optimisation program. This QCQP has been
studied before in the context of finding the ground state energy of a
particular Hamiltonian~\cite{bharti2020quantum}. We can recover the generalised
eigenvalue problem from the QCQP by introducing a Lagrange function
$L(\vec{v}, \lambda)$, and finding its stationary points
\begin{gather}
  L(\vec{v}, \lambda) = \vec{v}^\dag D \vec{v} + \lambda
  (1 - \vec{v}^\dag E \vec{v}) \\ 
  \frac{\partial L}{\partial \vec{v}} = 0 \implies D\vec{v} = \lambda E \vec{v}\,.
\end{gather}
From \eqref{eqn: H w.r.t CSK} and the observation that $\vec{v}_i^\dag E
\vec{v}_j = \delta_{ij}$ (see Appendix~\ref{Appendix: derivation of
diagonal representation}), we can write the evolution unitary within the
space spanned by the hybrid states as
\begin{equation}
  \label{eqn: fast forwarding}
  \left[e^{-i H T}\right]_{\mathbb{C}\mathbb{S}_K} = \sum_j e^{-i \lambda_j
  T} \vec{v}_j \vec{v}_j^\dag E\,.
\end{equation}
Using the notation that
$\left[\ket{\psi(\vec{\alpha}(T))}\right]_{\mathbb{C}\mathbb{S}_K}$ denotes
the coordinate vector of $\ket{\psi(\vec{\alpha}(T))}$ with respect to the
set of states $\mathbb{C}\mathbb{S}_K$, we see that the evolution of
$\vec{\alpha}(T)$ to a time $T$ is given by
\begin{align}
 \label{eqn: fast forwarding time evo}
  \vec{\alpha}(T)&=\left[\ket{\psi(\vec{\alpha}(T))}\right]_{\mathbb{C}\mathbb{S}_K} \nonumber \\ 
  &= \left[e^{-i H
  T}\ket{\psi(\vec{\alpha}(0))}\right]_{\mathbb{C}\mathbb{S}_K} = [e^{-i H
  T}]_{\mathbb{C}\mathbb{S}_K} \vec{\alpha}(0)\,.
\end{align}
\eqref{eqn: fast forwarding} is the reason why this method is called
"Classical-Quantum Fast Forwarding"; essentially what we are doing is
calculating the $D$ and $E$ matrices on the quantum computer, using those
matrices to find a diagonal representation of $H$ on a classical computer,
and finally using that to get $e^{-i H T}$ with a simple exponentiation of
the eigenvalues. After the fast forwarding, \eqref{eqn: fast forwarding time
evo} gives us $\vec{\alpha}(T)$ and hence $\ket{\psi(\vec{\alpha}(T))}$,
which is the expression for the time evolution under $H$ of
$\ket{\psi(\vec{\alpha}(t=0)}$ expressed as a linear combination of the
states in $\mathbb{C}\mathbb{S}_K$. The accuracy of our simulation can be
improved by increasing the value of $K$ in the definition of
$\mathbb{C}\mathbb{S}_K$.

We shall summarise the CQFF algorithm below as follows:
\begin{enumerate}
  \item{Get the hybrid ansatz state \eqref{eqn: hybrid ansatz equation} with
  classical parameters $\vec{\alpha}(t)$ and quantum states $\ket{\chi_i}$
  generated from an efficiently preferable state $\ket{\phi}$ and
  Hamiltonian $H$.}
  \item{Compute the $D$ and $E$ matrices on the quantum computer, with the
  matrix elements given in \eqref{eqn: D matrix element}~and~\eqref{eqn: E
  matrix element}.}
  \item{Solve the generalised eigenvalue problem in \eqref{eqn: generalised
  eigenvalue problem} and use \eqref{eqn: fast forwarding time evo} to get
  the time evolved parameters of the hybrid state $\vec{\alpha}(T)$ for some
  initial $\vec{\alpha}(0)$.}
\end{enumerate}

\section{Results}
We first use the CQFF algorithm to simulate the time evolution of the
Heisenberg model, given by the following Hamiltonian
\begin{equation}
  \label{eqn: XYZ Heisenberg model}
  H_1 = \sum_{j=1}^{N-1}  X_j X_{j+1} + 2 Y_j Y_{j+1} + 3 Z_j Z_{j+1}\,,
\end{equation}
where $N$ is the number of qubits. Here, we consider the $2$ and $3$ qubit
cases. Using IBM's  quantum processor \emph{ibmq\_rome}, we prepare a random
initial state $\ket{\phi}$ on the quantum computer (see appendix~\ref{Appendix:
initial state preparation} for more details). We also used the quantum computer
to calculate the matrix elements in \eqref{eqn: D matrix
element}~and~\eqref{eqn: E matrix element}, and the calculation of each matrix
element is done by just sampling the state $\ket{\phi}$ in some Pauli-rotated
basis. After obtaining the $D$ and $E$ matrices, the job of the quantum computer
is done; to finally obtain the time evolution of
$\ket{\psi(\vec{\alpha}(t=0)}=\ket{\phi}$, we use the classical computer to
perform the fast forwarding in accordance with \eqref{eqn: fast
forwarding}~and~\eqref{eqn: fast forwarding time evo}. To verify the results of
the time evolution, we compare the exact and experimental values found for
$\vec{\alpha}(T)$ by computing
\begin{equation}
  \ket{\psi(T)_{\text{theoretical}}} = e^{-iHT}\ket{\phi}
\end{equation}
classically as well as 
\begin{equation}
  \label{eqn: cqff state}
  \ket{\psi(T)_{\text{CQFF}}} = \sum_{i=1}^L\alpha_i(T)\ket{\chi_i}\,.
\end{equation}
We calculate fidelity over time given as
$F(t)=\left|\braket{\psi(t)_\text{theory}}{\psi(t)_\text{CQFF}}\right|^2$ and
the time variation of the expectation value $\langle Z_1 \rangle$, which we
denote as $\langle Z_1(t) \rangle$. It is easy to see that
we have
\begin{equation}
  \langle Z_1(t) \rangle = \sum_{i}^L \sum_{j}^L\alpha_i(t)^*
  \mel{\chi_i}{Z_1}{\chi_j} \alpha_j(t).
\end{equation}
Here, we focus on the ability of our algorithm to accurately obtain the
coefficients of the time evolution $\vec{\alpha}(T)$ in \eqref{eqn: cqff
state} using the quantum computer.  While $\vec{\alpha}(T)$ is calculated
using the $D$ and $E$ matrices that were computed on the quantum computer,
the overlap terms for the expectation values $\mel{\chi_i}{Z_1}{\chi_j}$
are computed classically. For completeness, we also show the $2$ qubit case
in Appendix~\ref{Appendix: Supplementary plots} where terms like
$\mel{\chi_i}{Z_1}{\chi_j}$ are computed on the
quantum computer as well.
\begin{figure}
  \centering
  \includegraphics[width=0.27\textwidth, trim={0cm 0cm 0cm 0cm},clip]{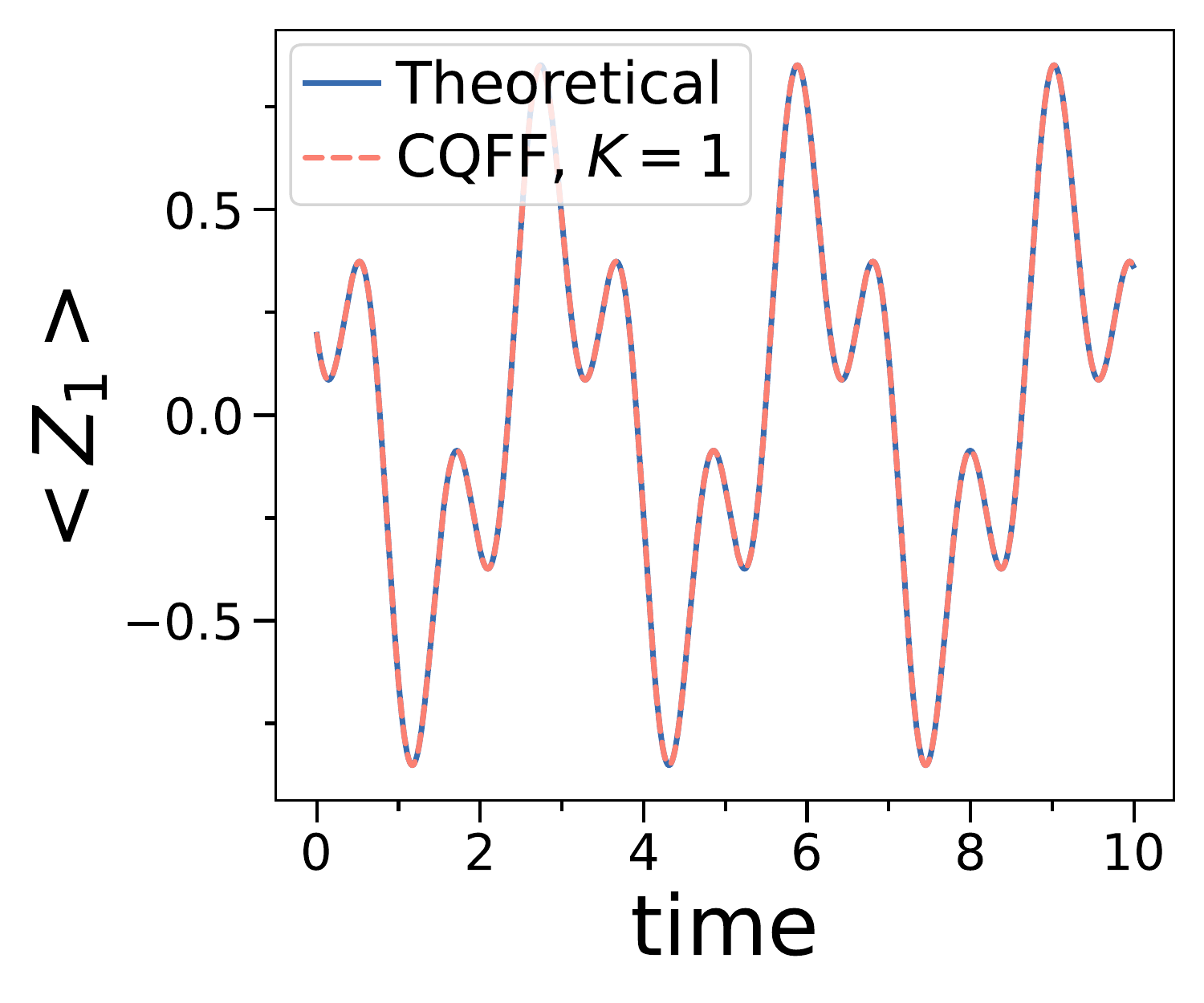}
  \caption[]{Time evolution of CQFF for 2 qubits with Hamiltonian
  $H_1$, simulated on the IBM quantum processor \emph{ibmq\_rome} with $8192$
  shots. The expectation value $\langle Z_1 \rangle$ is shown here. The
  fidelity of the state remains $1$ for the entire time evolution and is
  hence omitted here. We plot the long time
  behavior of the fidelity in Appendix~\ref{Appendix: Supplementary
  plots}.}
  \label{fig: 2 qubit XYZ results}
\end{figure}
\begin{figure}
  \centering
  \subfigimg[width=0.24\textwidth]{a}{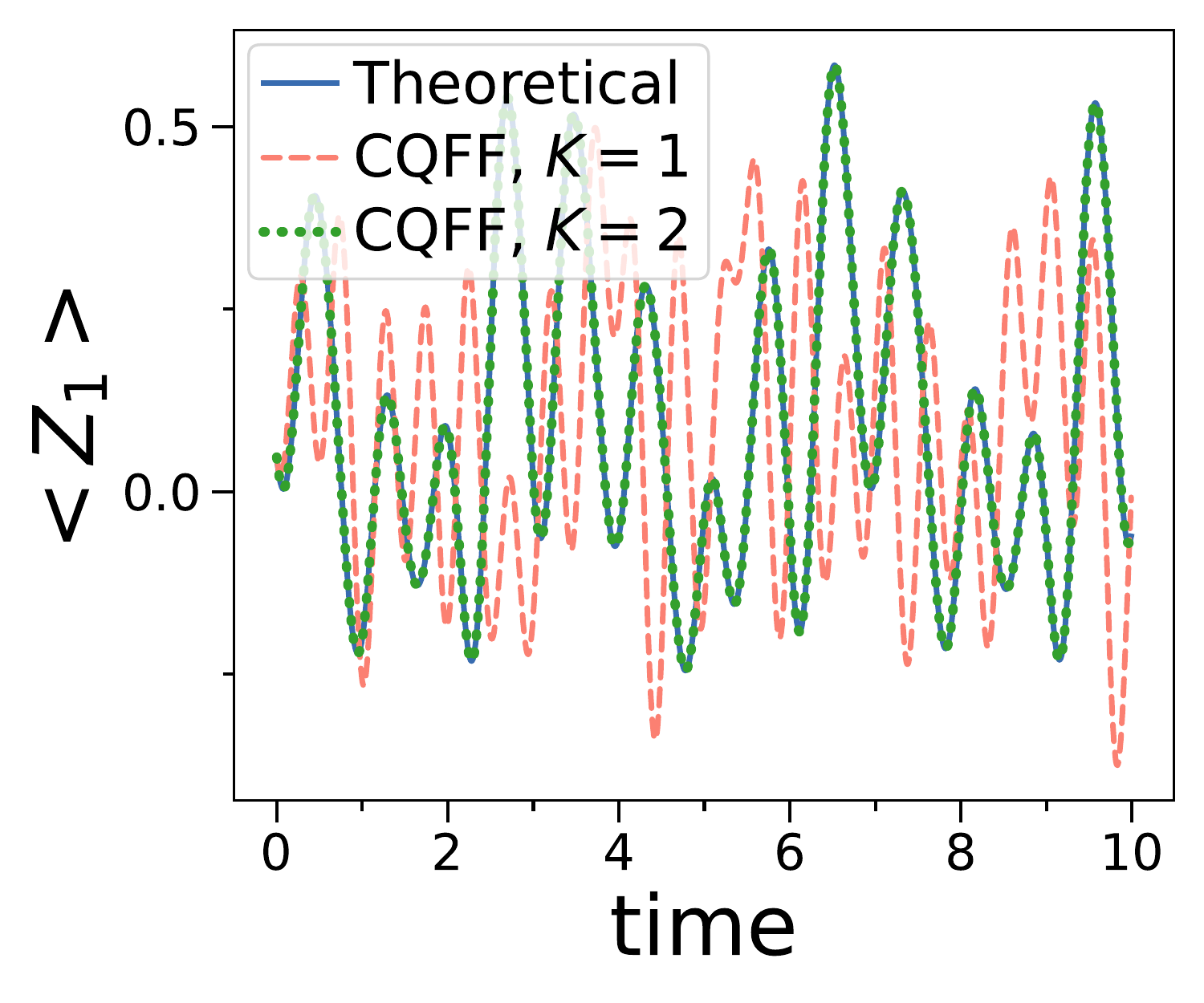}\hfill
  \subfigimg[width=0.24\textwidth]{b}{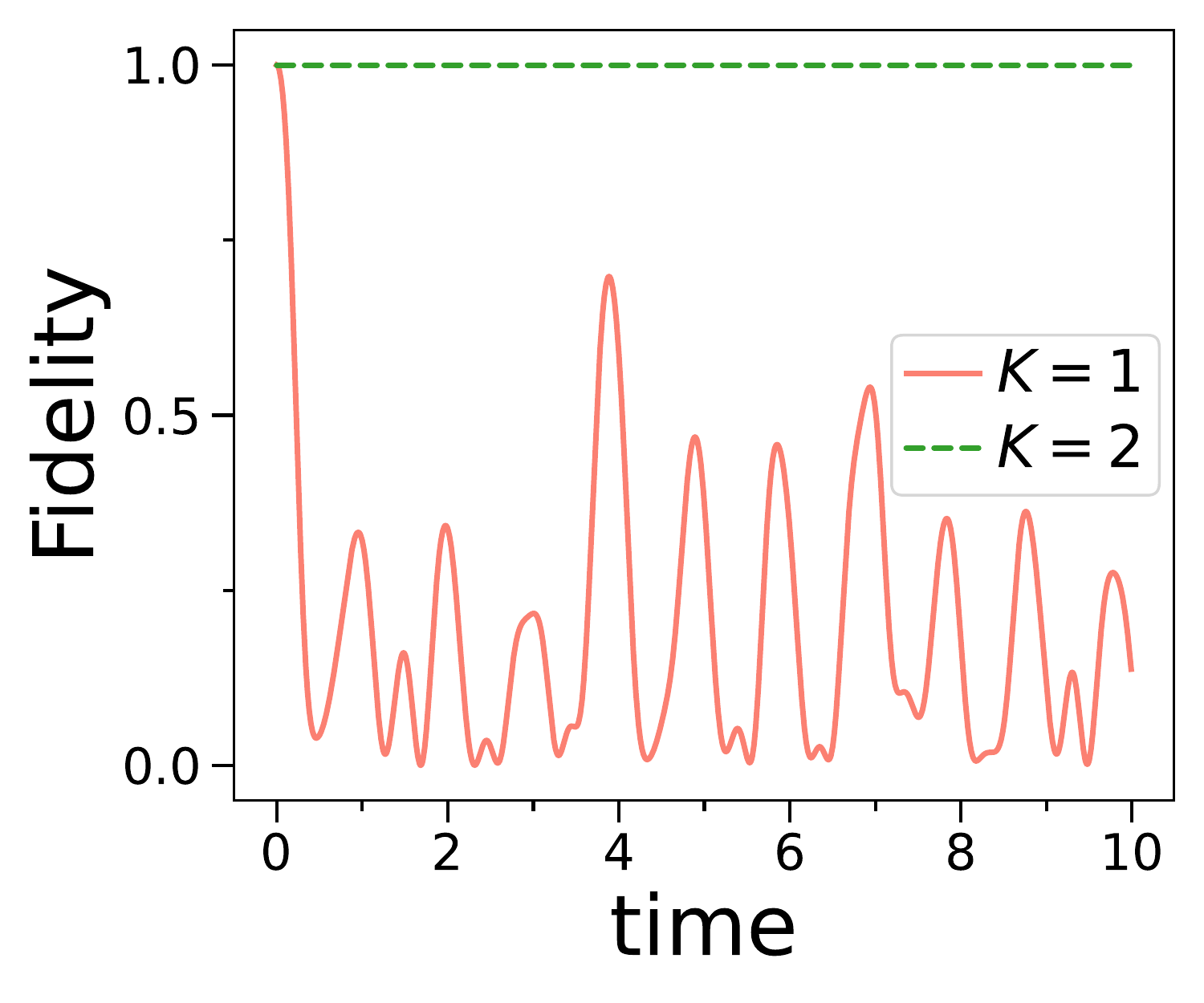}
  \caption[]{Time evolution of CQFF on a 3 qubit state with Hamiltonian
  $H_1$, simulated on the IBM quantum processor \emph{ibmq\_rome} with $8192$
  shots. \textbf{a)} Expectation value $\langle Z_1 \rangle$.
  The CQFF $K=2$ values completely match with the
  exact values. \textbf{b)} Fidelity of the state.
  We also plot the long time behavior of the fidelity
  in Appendix~\ref{Appendix: Supplementary plots}}
  \label{fig: 3 qubit XYZ results}
\end{figure}
The results for both the fidelity $F(t)$ and for $\langle Z_1(t) \rangle$ are
shown in \figref{fig: 2 qubit XYZ results} and~\figref{fig: 3 qubit XYZ
results} for the $2$ and $3$ qubit cases.

\begin{figure}
  \centering
  \subfigimg[width=0.24\textwidth]{a}{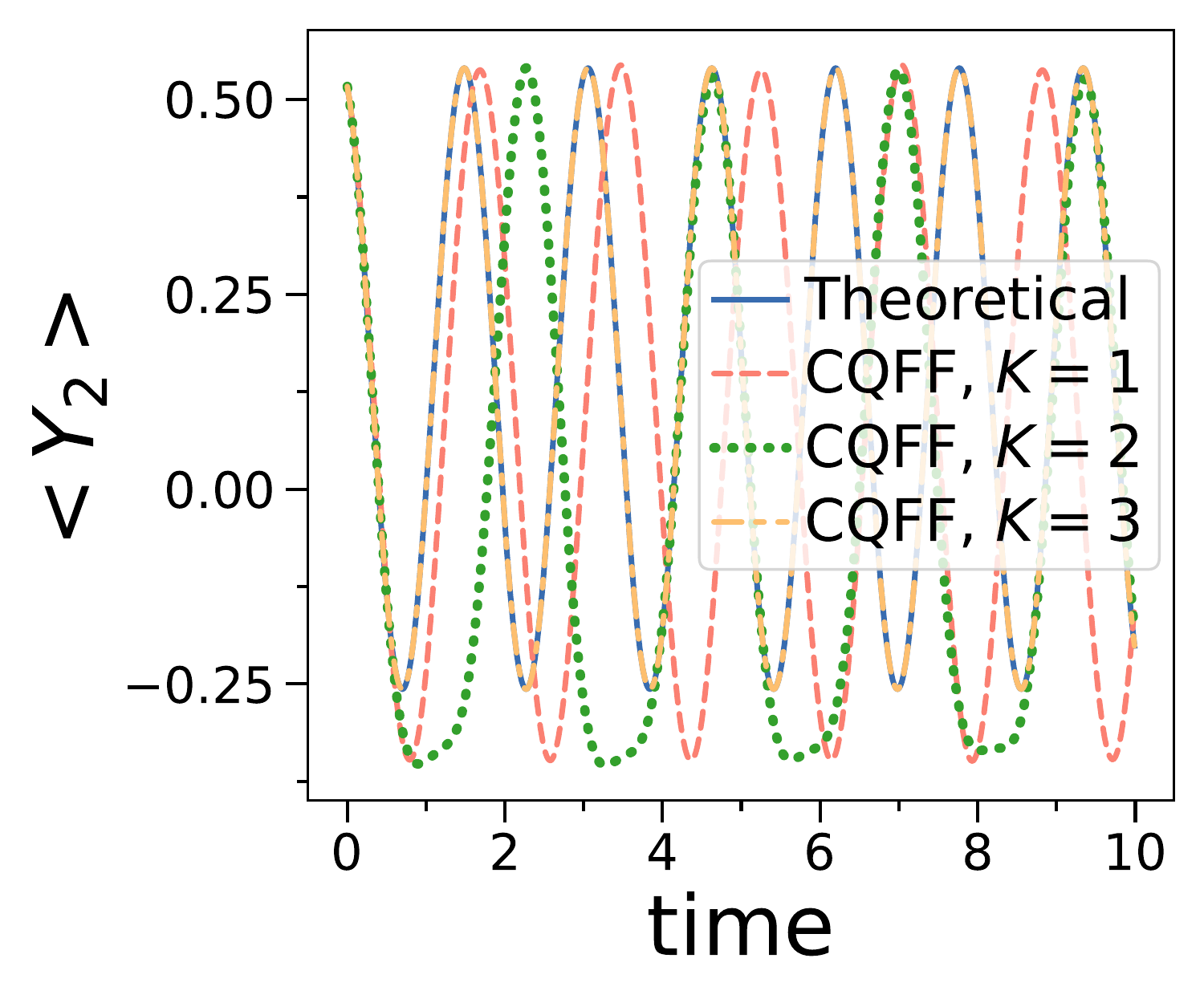}\hfill
  \subfigimg[width=0.24\textwidth]{b}{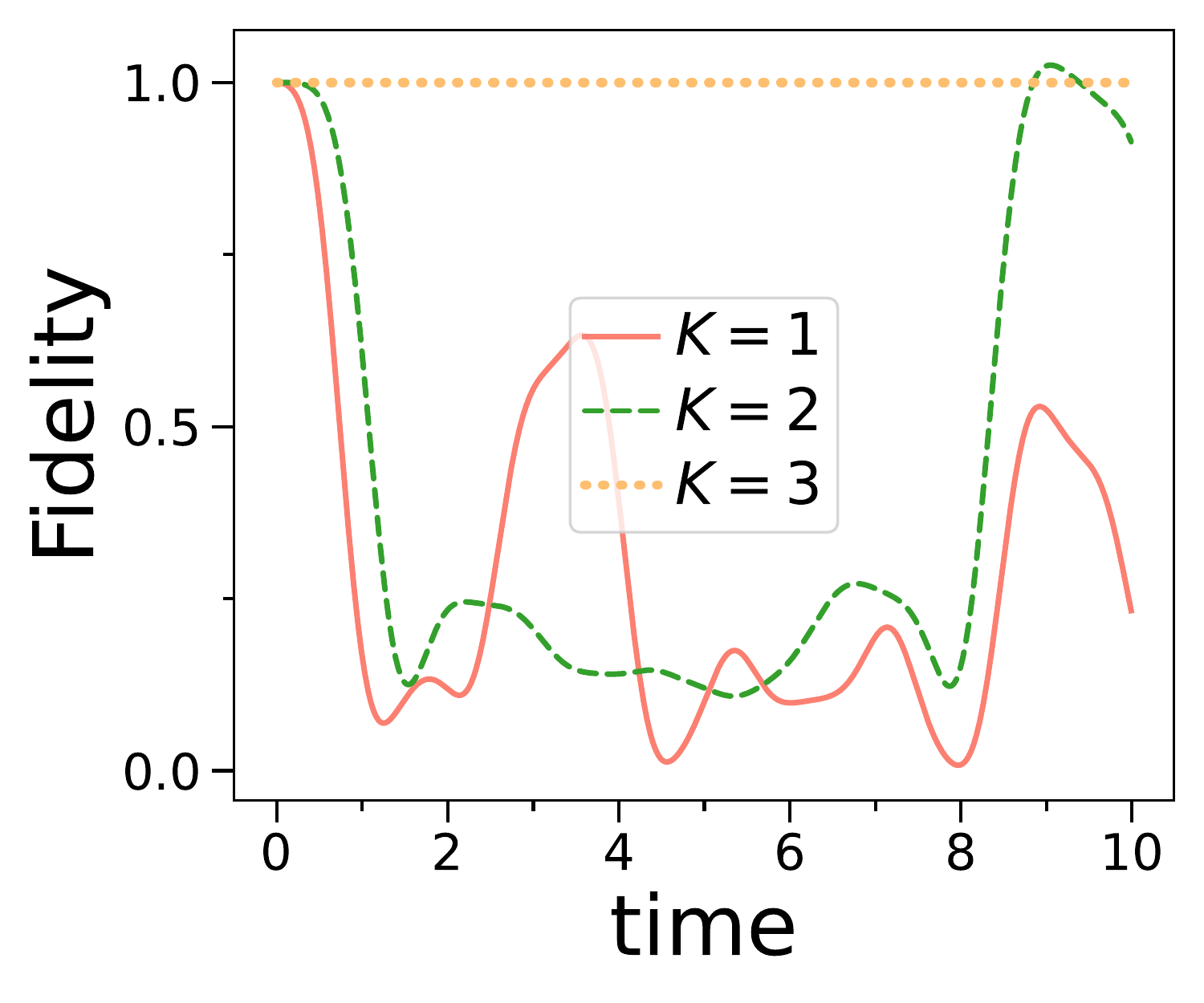}
  \caption[]{Time evolution of CQFF on a 5 qubit state with Hamiltonian
  $H_2$, simulated on the IBM quantum processor \emph{ibmq\_rome} with $8192$
  shots. Here, we have $J_{zxz} = 1$. \textbf{a)} Expectation value $\langle
  Y_2 \rangle$ \textbf{b)} Fidelity of the state. We
  plot the long time behavior of the fidelity in Appendix~\ref{Appendix:
  Supplementary plots}.}
  \label{fig: 5 qubit three-body results}
\end{figure}

Next, we use the CQFF algorithm to simulate a three-body Hamiltonian proposed
in \cite{petiziol2020quantum}
\begin{equation}
  H_2 = J_{zxz}\sum_{k=1}^{N} Z_{k-1}X_kZ_{k+1}\,.
\end{equation}
We consider the $4$ and $5$ qubit cases, and follow the same procedure as for
the Heisenberg model above on the same quantum computer \emph{ibmq\_rome}.
The random initial state $\ket{\phi}$ is also prepared the same way. To
verify our results, we compute the time variation of the fidelity $F(t)$ as
well as the time variation of the expectation value $\langle Y_2 \rangle$,
which we denote as $\langle Y_2(t) \rangle$. 
As before, coefficients $\vec{\alpha}(T)$ are calculated
using the $D$ and $E$ matrices that were computed with the quantum computer,
while the terms $\mel{\chi_i}{Y_2}{\chi_j}$ needed for the expectation values
are calculated using classical computers.
The results are shown in \figref{fig: 4 qubit three-body
results}~and~\figref{fig: 5 qubit three-body results}.
\begin{figure}
  \centering
  \subfigimg[width=0.24\textwidth]{a}{ 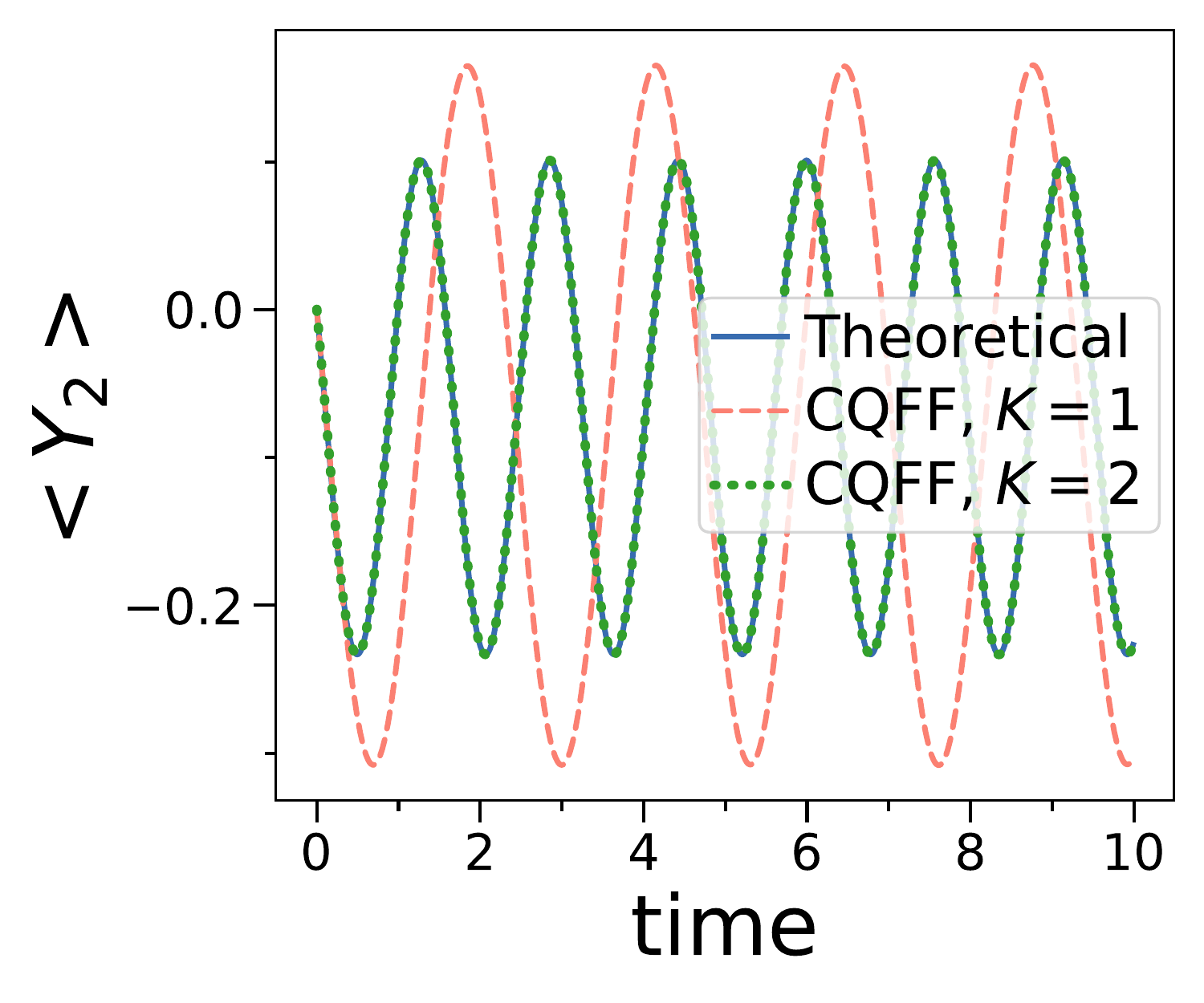}\hfill
  \subfigimg[width=0.24\textwidth]{b}{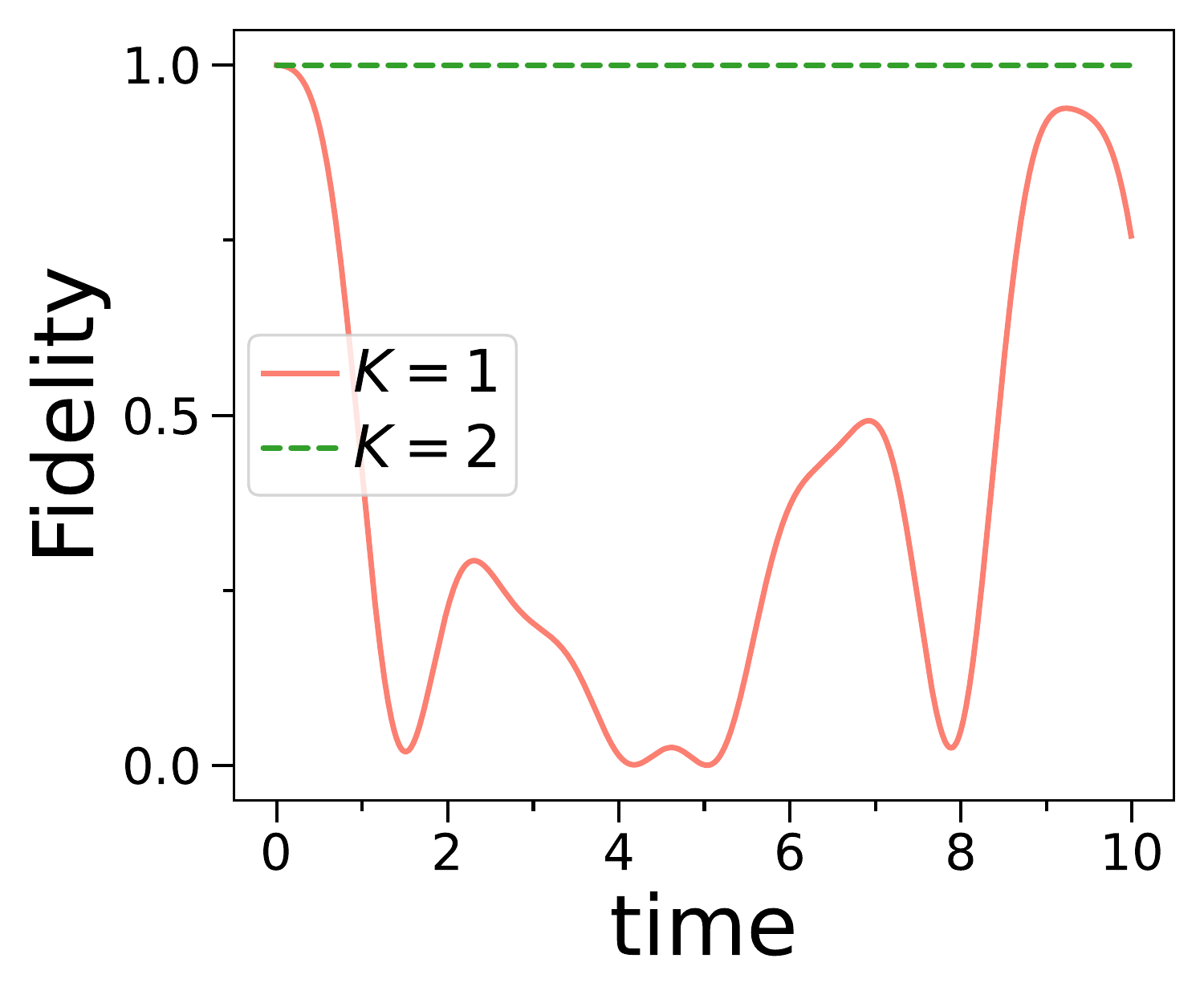}
  \caption[]{Time evolution of CQFF on a 4 qubit state with Hamiltonian
  $H_2$, simulated on the IBM quantum processor \emph{ibmq\_rome} with $8192$
  shots. Here, we have $J_{zxz} = 1$. \textbf{a)} Expectation value of
  $\langle Y_2 \rangle$. The CQFF $K=2$ values completely
  match with the exact values. \textbf{b)} Fidelity of the state.
  We also plot the long time behavior of the fidelity
  in Appendix~\ref{Appendix: Supplementary plots}.}
  \label{fig: 4 qubit three-body results}
\end{figure}
We see that for both the Hamiltonians $H_1$ and $H_2$ considered, we could
essentially perform quantum simulation for up to $t=10$ by considering a
large enough value for $K$.

\begin{figure}
  \centering
  \subfigimg[width=0.24\textwidth]{a}{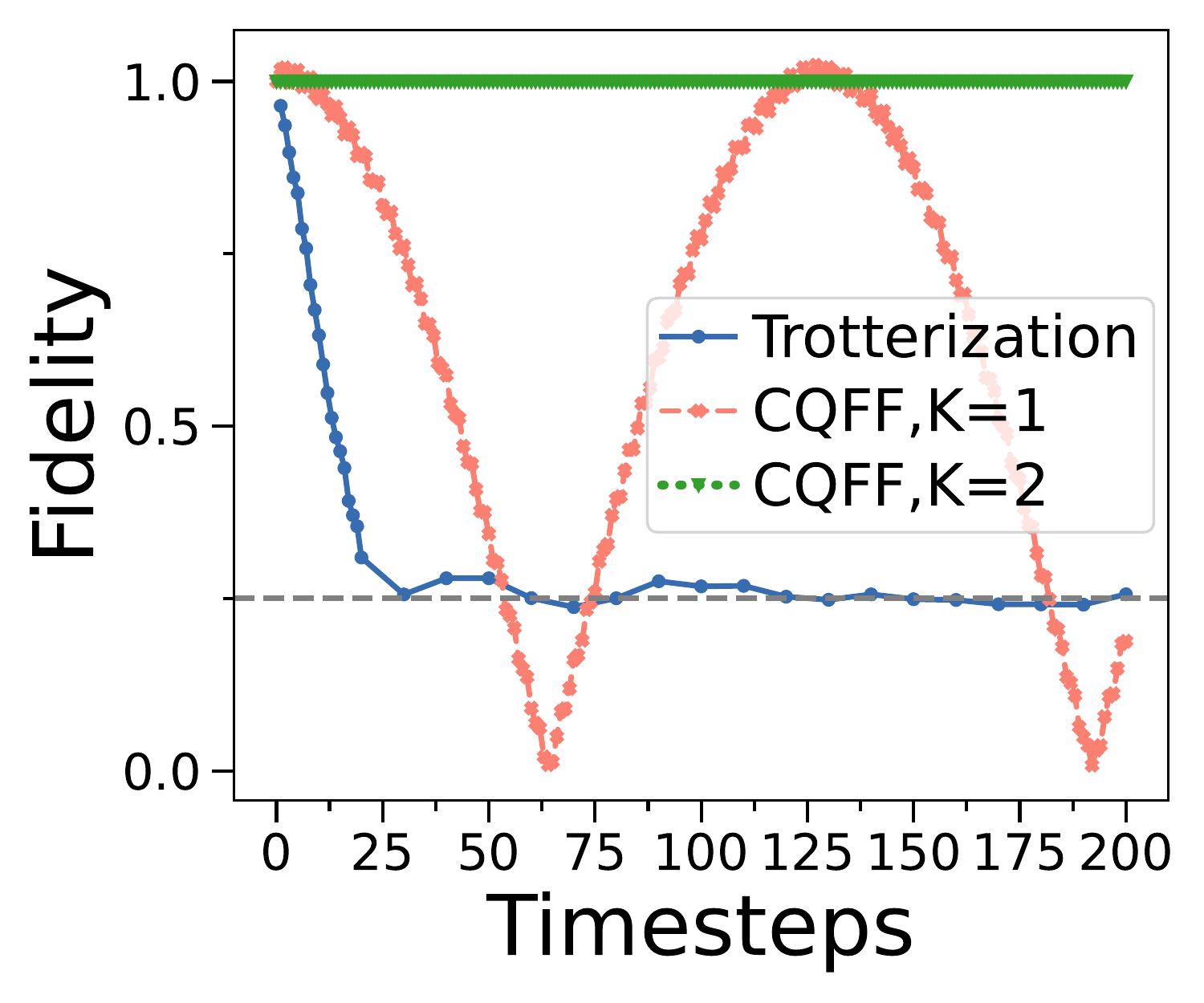}\hfill
  \subfigimg[width=0.24\textwidth]{b}{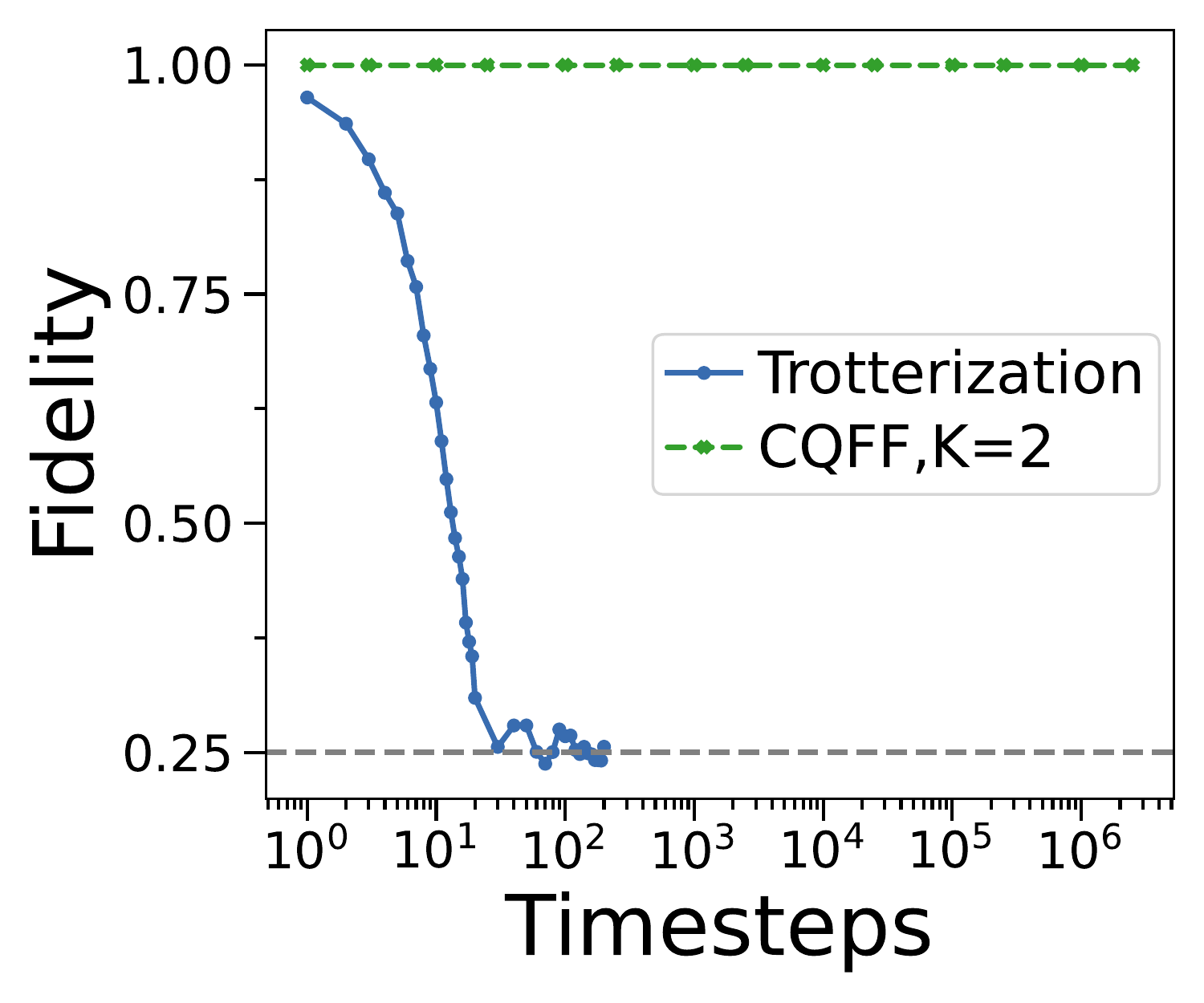}
  \caption[]{Time evolution of both CQFF and Trotterization for the $2$-qubit
  state $\ket{10}$ under the Hamiltonian $H_3$, simulated on the IBM quantum
  processor \emph{imbq\_rome} with $8192$ shots. Here, we used the first-order
  Trotter approximation with a timestep $\Delta t = 0.5$. The grey dashed line
  at $F = 0.25$ represents the overlap with the maximally mixed state.
  \textbf{a)} Plotted on a linear scale, up to $200$ time steps \textbf{b)}
  Plotted on a log scale, up to $2500000$ timesteps.  The fidelity for CQFF
  $K=1$ is still oscillatory even at long time scales and is hence omitted in
  the log plot.}
  \label{fig: fidelity comparison plot}
\end{figure}

Next, we benchmark the CQFF algorithm against Trotterization on the IBM quantum
processor \emph{ibmq\_rome}. To date, the best known fast forwarding on real
quantum hardware has been \cite{gibbs_long-time_2021}, which managed to maintain
a fidelity of at least $0.9$ for at least $600$ Trotter steps. Here, we show
that we can surpass this by a factor of about $10^4$. Like
\cite{gibbs_long-time_2021}, we consider the time-evolution of the state
$\ket{\psi(t=0)} = \ket{10}$ under the $2$-qubit $XY$ spin chain Hamiltonian
given by
\begin{equation}
  H_{3} = X_1 X_2 + Y_1Y_2 \,.
\end{equation}
We time evolve the state $\ket{\psi(t=0)}$ using both Trotterization and CQFF,
where for Trotterization, we consider the first-order Trotter approximation with
a timestep $\Delta t = 0.5$. We evaluate the quality of our simulations by
plotting the fidelity $F = \mel{\psi(t)}{\rho(t)}{\psi(t)}$ against the number
of Trotter steps $N$, where $\ket{\psi(t)}$ here is the exact evolution and
$\rho(t)$ is the simulated evolution. For Trotterization the fidelity is
computed by first performing tomography, whereas since CQFF uses the hybrid
ansatz in \eqref{eqn: hybrid ansatz equation}, there is no need for tomography.
The results are shown in \figref{fig: fidelity comparison plot}.  The grey
dashed line at $F = 0.25$ represents the fidelity with the maximally mixed
state. As can be seen, Trotterization breaks down at around $25$ Trotter steps
due to decoherence. On the other hand, for the entire period of the time
evolution, CQFF for $K = 1$ exhibits $F$ that is sinusoidally varying between
$0$ and slightly greater than $1$, and CQFF for $K = 2$ has a fidelity of $1$.
The fidelity being slightly greater than 1 is due to numerical errors in the
$E$-matrix. As seen in the log plot of \figref{fig: fidelity comparison plot},
the same behavior for $K = 2$ continues even on a longer timescale, at $2500000$
Trotter steps.  Hence, we can conclude that on current term quantum hardware,
the CQFF algorithm with $ K = 2$ allows for a fast-forwarding of about $10^5$
times compared to the coherence time of the quantum computer, which is
essentially indefinitely long.

\section{Discussion}
The accuracy of the CQFF algorithm increases as we consider larger $K$ in
$\mathbb{C}\mathbb{S}_K$.  As $K$ increases, the set $\mathbb{C}\mathbb{S}_K$
is able to better span the space of states necessary for time evolution of
the initial state under $H$.  Furthermore, the quantum computer can calculate
the $D$ and $E$ matrices in a single step without any feedback loop.
This also means that any noise due to the computation on
the quantum computer only affects our algorithm in this single step.  More
information about error analysis is given in Appendix~\ref{Appendix: Error
analysis}.

For the examples considered above, we just needed to sample the state
$\ket{\phi}$ in some Pauli-rotated basis, which can be done efficiently on
NISQ quantum devices in contrast to more complicated routines like the
Hadamard test which requires controlled multi-qubit unitaries.  The barren
plateau problem is also avoided by construction, as there is no parameterized
quantum circuit that is being updated.

For the CQFF algorithm, further studies on the scaling of number of states in
$\mathbb{C}\mathbb{S}_K$ with the number of qubits need to be performed. The
rate of growth in the number of states in $\mathbb{C}\mathbb{S}_K$ as the
number of qubits increases depends highly on the Hamiltonian; for example,
$H_2$ has lesser number of states in $\mathbb{C}\mathbb{S}_K$ as compared to
$H_1$ for the same number of qubits. Here, we propose that since we are using
hybrid states for our ansatz, as defined in \eqref{eqn: hybrid ansatz
equation}, whether a Hamiltonian can be fast-forwarded with the CQFF
algorithm depends on the growth of the number of states in
$\mathbb{C}\mathbb{S}_K$ with the number of qubits. It remains an open
question to study which Hamiltonians can be fast-forwarded. More work can
also be done to study how one might reduce the states in
$\mathbb{C}\mathbb{S}_K$ for a given Hamiltonian $H$. Lastly, more work has
to be done to study how the choice of the state $\ket{\phi}$ affects the
scaling and performance of the CQFF algorithm.

We note that CQFF as formulated above is more similar to VHD than VFF.  One
can also tweak the CQFF algorithm to make it more similar to VFF.  This can
be done by using CQFF to diagonalise a small $\Delta t$ approximation of
$U(\Delta t) = e^{-i H \Delta t}$ instead of $H$.  Then, with the diagonal
representation of $U(\Delta t)$, we can easily get $U(N\Delta t)$ (See
Appendix~\ref{Appendix: tweaking CQFF to look like VFF} for more details).
However, this has no advantages in our framework and is instead
disadvantageous as the small $\Delta t$ approximation introduces errors that
are avoided by directly diagonalising $H$ as per the original CQFF algorithm.

The CQFF algorithm can also make use of a classical quantum feedback loop if
we allow for the quantum states $\ket{\chi_i}$ defining our hybrid ansatz in
\eqref{eqn: hybrid ansatz equation} to be variationally adjusted.  In other
words, we can define a hybrid ansatz like
\begin{equation}
    \ket{\psi(\vec{\theta},\vec{\alpha}(t))} = \sum_{i=1}^L \alpha_i(t) \ket{\chi_i(\vec{\theta})}
\end{equation}
where $\ket{\chi_i(\vec{\theta})} = U_i \ket{\phi(\vec{\theta})}$ and $U_i$
here is a product of unitaries in $\mathbb{U}$.  Using this idea, we can
control our state $\ket{\psi(\vec{\theta},\vec{\alpha}(t))}$ both
variationally by updating $\vec{\theta}$ and classically by updating
$\vec{\alpha}(t)$.

\appendix
\section{Justification of CQFF algorithm}
\label{Appendix: derivation of diagonal representation}
Here, we first want to find the representation matrix of $H$ with respect to
the set $\mathbb{C}\mathbb{S}_K$, which we define as
$[H]_{\mathbb{C}\mathbb{S}_K}$. $[H]_{\mathbb{C}\mathbb{S}_K}$ is given by
the following equation:
\begin{equation}
 \label{eqn: LA representation matrix theorem}
 [H]_{\mathbb{C}\mathbb{S}_K} = \begin{pmatrix} \vec{u}_1 \,\,\, \vec{u}_2 \dots \vec{u}_L \end{pmatrix}
\end{equation}
where $\vec{u}_i = [H\ket{\chi_i}]_{\mathbb{C}\mathbb{S}_K}$ is the
coordinate vector of $H\ket{\chi_i}$ with respect to the set
$\mathbb{C}\mathbb{S}_K$. From \eqref{eqn: LA representation matrix theorem},
we see that if we define:
\begin{equation}
  \label{eqn: defn of B}
  B = \begin{pmatrix}
  \ket{\chi_1} \,\, \ket{\chi_2} \,\, \dots \,\, \ket{\chi_L}
\end{pmatrix}
\end{equation}
we have:
\begin{align}
  HB 
  &= H\begin{pmatrix} \ket{\chi_1} \,\, \ket{\chi_2} \,\, \dots \,\,
  \ket{\chi_L} \end{pmatrix} \nonumber \\
  &= \begin{pmatrix} H\ket{\chi_1} \,\, H\ket{\chi_2} \,\, \dots \,\,
  H\ket{\chi_L} \end{pmatrix} \nonumber \\
  \label{eqn: HB = B[H] eqn}
  &= B[H]_{\mathbb{C}\mathbb{S}_K} \,.
\end{align}
Note that defining $B$ in \eqref{eqn: defn of B} also leads us to two very
helpful expressions:
\begin{align}
  B^\dag B = E \\
  B^\dag H B = D
\end{align}
where $B^\dag$ is: 
\begin{equation}
  B^\dag = \begin{pmatrix}
  \bra{\chi_1} \\ \bra{\chi_2} \\ \vdots \\ \bra{\chi_L} \end{pmatrix}\,.
\end{equation}
Hence if we multiply $B^\dag$ from the left to both sides of \eqref{eqn: HB =
B[H] eqn}, we will get:
\begin{align}
  \label{eqn: D = E[H]}
  D = E[H]_{\mathbb{C}\mathbb{S}_K}\,.
\end{align}

We will show next that $[H]_{\mathbb{C}\mathbb{S}_K}$ can be related to the
eigenvectors $\vec{v}_i$ obtained from solving the generalised
eigenvalue problem in \eqref{eqn: generalised eigenvalue problem}.

When solving the generalised eigenvalue problem in \eqref{eqn: generalised
eigenvalue problem}, we note that $E$ in general is not full rank and hence
some of the eigenvectors $\{\vec{v}_i\}$ obtained will belong to the
nullspace of $E$. Since $D$ is a Hermitian matrix and $E$ is a positive
semi-definite matrix, the eigenvectors $\vec{v}_i$ that are in the
column space of $E$ can always be chosen to be orthonormal with respect to
$E$. In other words, we can always have:
\begin{align}
  \label{eqn: orthonormality of alpha vec}
  \vec{v}_i^\dag E \vec{v}_j &= \delta_{ij} \,\, \text{if } \vec{v}_i\text{
  and }\vec{v}_j \in \text{Col}(E) \\
  \vec{v}_i^\dag E \vec{v}_j &= 0 \,\, \text{if } \vec{v}_i\text{ or
  }\vec{v}_j \in \text{Null}(E)
\end{align}
where $\text{Col}(E)$ stands for the column space of $E$, $\text{Null}(E)$
stands for the nullspace of $E$, and $\delta_{ij}$ here is the
Kronecker-Delta symbol. Now, consider the expression:
\begin{equation}
  \sum_{k} \vec{v}_i^\dag E \vec{v}_k \vec{v}_k^\dag E
  \vec{v}_j 
  = \vec{v}_i^\dag E \left(\sum_{k} \vec{v}_k
  \vec{v}_k^\dag E\right) \vec{v}_j\,.
\end{equation}
If we then choose $\vec{v}_i\text{ and }\vec{v}_j \in
\text{Col}(E)$, we have:
\begin{align}
  \vec{v}_i^\dag E \left(\sum_{k} \vec{v}_k
  \vec{v}_k^\dag E\right) \vec{v}_j
  = \sum_{k} \delta_{ik} \delta_{kj} = \delta_{ij}
\end{align}
which then immediately leads to the following completeness relation:
\begin{equation}
  \label{eqn: completeness eqn}
  \sum_{j} \vec{v}_j\vec{v}_j^\dag E = \mathbbm{1}\,.
\end{equation}
Note that the expression $\left(\sum_{k} \vec{v}_k \vec{v}_k^\dag
E\right)$ is the same regardless of whether we include the vectors
$\vec{v}_k \in \text{Null}(E)$. Hence from now on, we will ignore these
nullspace eigenvectors in \eqref{eqn: completeness eqn}.

Now, what we do is to multiply the $D$ matrix from the left in
\eqref{eqn: completeness eqn} to get
\begin{align}
  &D = D\sum_{j} \vec{v}_j\vec{v}_j^\dag E \nonumber \\
  \label{eqn: D = E H_spectral_decomp_csk}
  \implies& D = E \left(\sum_{j} \lambda_j \vec{v}_j\vec{v}_j^\dag E\right)\,.
\end{align}
Direct comparison of \eqref{eqn: D = E H_spectral_decomp_csk} with
\eqref{eqn: D = E[H]} gives us
\begin{equation}
  \label{eqn: H spectral_decomp_csk}
  [H]_{\mathbb{C}\mathbb{S}_K} = \sum_i \lambda_i
  \vec{v}_i\vec{v}_i^\dag E + N\,,
\end{equation}
where $N$ is a matrix whose columns are in the nullspace of $E$. We can then
use \eqref{eqn: H spectral_decomp_csk}~and~\eqref{eqn: orthonormality
of alpha vec} to arrive at
\begin{equation}
  \left[e^{-iHT}\right]_{\mathbb{C}\mathbb{S}_K} = \sum_i e^{-i \lambda_iT}
  \vec{v}_i\vec{v}_i^\dag E + N^{\prime}\,,
\end{equation}
where $N'$ is another matrix whose columns are in the nullspace of $E$. Now,
for the initial state $\ket{\psi(\vec{\alpha}(0))}$ to be a valid quantum
state, it must be normalised, which corresponds to the condition
$\vec{\alpha}(0)^\dag E \vec{\alpha}(0) = 1$. I.e, $\vec{\alpha}(0)$ must be
in the column space of $E$. This means that
\begin{align}
  \left[e^{-iHT}\right]_{\mathbb{C}\mathbb{S}_K}\vec{\alpha}(0)
  &= \left(\sum_i e^{-i \lambda_iT} \vec{v}_i\vec{v}_i^\dag
  E\right)\vec{\alpha}(0) + N^{\prime}\vec{\alpha}(0) \nonumber \\
  \label{eqn: fast-forwarding time evo with nullspace vec}
  &= \left(\sum_i e^{-i \lambda_iT} \vec{v}_i\vec{v}_i^\dag
  E\right)\vec{\alpha}(0) + \vec{n}\,,
\end{align}
where $\vec{n}$ is a vector in the nullspace of $E$. Now, we can safely
ignore the vector $\vec{n}$, since it corresponds to the zero ket in the
Hilbert space; to see why, we write
\begin{equation}
  \ket{\vec{n}} = \sum_i n_i \ket{\chi_i}
\end{equation}
and compute $\braket{\vec{n}}{\vec{n}}$:
\begin{equation}
  \braket{\vec{n}}{\vec{n}} = \vec{n}^\dag E \vec{n} = 0\,.
\end{equation}
Since $\braket{\vec{n}}{\vec{n}} = 0$, $\ket{\vec{n}}$ must be the zero ket,
and hence we can ignore $\vec{n}$ in \eqref{eqn: fast-forwarding
time evo with nullspace vec}. In the end, we arrive at:
\begin{equation}
  \left[e^{-iHT}\right]_{\mathbb{C}\mathbb{S}_K}\vec{\alpha}(0)
  = \left(\sum_i e^{-i \lambda_iT} \vec{v}_i\vec{v}_i^\dag
  E\right)\vec{\alpha}(0)
\end{equation}
which is exactly the CQFF time evolution equation given in \eqref{eqn: fast forwarding time evo}.

\section{Cumulative K-moment states}
\label{Appendix: CSK definition}
Here, we first formally define the $\mathbb{C}\mathbb{S}_K$ states before we
explain how they can capture Hamiltonian time evolution. The definition of
$\mathbb{C}\mathbb{S}_K$ here is taken from \cite{bharti2020iterative}. 
\begin{definition} (Adapted from \cite{bharti2020iterative}.)
  Given a set of unitaries $\mathbb{U} \equiv \{U_i\}_{i=1}^r$, a positive
  integer $K$ and some quantum state $\ket{\phi}$, the $K$-moment states is
  the set of quantum states of the form $\{U_{i_K}\dots U_{i_2}
  U_{i_1}\ket{\phi}\}_i$ for $U_{il} \in \mathbb{U}$. We denote the
  aforementioned set by $\mathbb{S}_K$. The singleton set
  $\{\ket{\phi}\}$ will be referred to as the $0$-moment state (denoted by
  $\mathbb{S}_0$). The cumulative $K$-moment states $\mathbb{C}\mathbb{S}_K$
  is defined to be $\mathbb{C}\mathbb{S}_K \equiv \cup_{j=0}^K \mathbb{S}_j$.
\end{definition}
In this paper, if the Hamiltonian $H$ is given by \eqref{eqn: Hamiltonian as
linear combination of unitaries}, we shall let $\mathbb{U}$ be the set of
unitaries that make up our Hamiltonian $H$, and we will use $\mathbb{U}$ to
construct $\mathbb{C}\mathbb{S}_K$. We then have:
\begin{align*}
  \mathbb{C}\mathbb{S}_0 &= \mathbb{S}_0  \\
  &= \{\ket{\phi}\} \\
  \mathbb{C}\mathbb{S}_1 &= \mathbb{C}\mathbb{S}_0 \cup \mathbb{S}_1 \\
  &= \{\ket{\phi}\} \cup \{U_{i_1} \ket{\phi}\}_{i_1=1}^r, \\
  \mathbb{C}\mathbb{S}_2 &= \mathbb{C}\mathbb{S}_1 \cup \mathbb{S}_2 \\
  &= \{\ket{\phi}\} \cup \{U_{i_1} \ket{\phi}\}_{i_1=1}^r \cup \{U_{i_2}
  U_{i_1} \ket{\phi}\}_{i_1=1, i_2 = 1}^r \\
  \vdots \\
  \mathbb{C}\mathbb{S}_K &= \mathbb{C}\mathbb{S}_{K-1} \cup \mathbb{S}_K
\end{align*}
where, $U_{i_l} \in \mathbb{U}$. 

The motivation for using the set $\mathbb{C}\mathbb{S}_K$ to construct our
ansatz can be found in \cite{bharti2020simulator}, but we will reproduce the
main gist of their argument here. Starting with $\ket{\psi(t=0)}=\ket{\phi}$,
the time evolution $\ket{\psi(t)} = e^{-iHt}\ket{\psi(t=0)}$ can be approximated as 
\begin{equation}
  \label{eqn: krylov subspace approx}
  \ket{\psi(t)} = e^{-iHt}\ket{\phi} \approx p_{m-1}(-iHt)\ket{\phi},
\end{equation}
where $p_{m-1}$ is some $m-1$ degree polynomial\cite{saad1992analysis}. Now,
$p_{m-1}(-iHt)\ket{\phi}$ is an element of the Krylov subspace
$\mathcal{K}_{m}$, which is defined as
\begin{equation}
  \label{eqn: krylov subspace definition (appendix)}
  \mathcal{K}_m = \text{span}\{\ket{\phi}, H\ket{\phi}, \dots, H^{m-1}
  \ket{\phi}\}.
\end{equation}
Hence, the problem of Hamiltonian time evolution up to a time $t$ can be
reframed as the problem of finding a linear combination of vectors in
$\mathcal{K}_m$, where the coefficients of that linear combination are
functions of $t$. We note that if the Hamiltonian $H$ has rank $r$, then the
approximation in \eqref{eqn: krylov subspace approx} becomes exact for some
$m \leq r+1$. Now, if we assume that our Hamiltonian $H$ is a linear
combination of unitaries as per \eqref{eqn: Hamiltonian as linear combination
of unitaries}, then it is easy to see that
\begin{equation}
  \mathcal{K}_{K+1} \subset \mathbb{C}\mathbb{S}_K
\end{equation}
and hence by using a linear combination of states in $\mathbb{C}\mathbb{S}_K$
for our ansatz, we leverage on the power of the Krylov subspace $K_{K+1}$ to
approximate the time evolution $e^{-iHt}\ket{\phi}$.

If the unitaries in our Hamiltonian $H$ are tensored-Pauli operators, then
the set $\mathbb{C}\mathbb{S}_K$ has a lot more mathematical structure that
can be exploited, due to the Lie Algebra structure of the tensored-Pauli
operators. This means that in some cases, we need to calculate a lot less
overlaps on the quantum computer than what is initially suggested by the
construction of the set $\mathbb{C}\mathbb{S}_K$.

\section{Error analysis}
\label{Appendix: Error analysis}
Due to the lack of the classical-quantum feedback loop in the CQFF algorithm,
any error in the CQFF algorithm for a given value of $K$ can be traced back
to the computation of the matrix elements $E_{ij} = \braket{\chi_i}{\chi_j}$
and $D_{ij} = \mel{\chi_i}{H}{\chi_j}$ on the quantum computer, where
$\ket{\chi_i},\ket{\chi_j} \in \mathbb{C}\mathbb{S}_K$. In this section we
first discuss the errors in the computation of those matrix elements before
discussing how these errors are propagated in our algorithm.

Firstly, the sources of error involved in the computation of $E_{ij}$ and
$D_{ij}$ on the quantum computer are:
\begin{enumerate}
  \item{Shot noise error due to a finite number of shots.}
  \item{Initial state preparation error in the preparation of $\ket{\phi}$.}
  \item{Final measurement error.}
\end{enumerate}

Since the Hamiltonians considered in this paper are tensored-Pauli operators,
which we denote by $P_k$, the matrix elements computed on the quantum
computer can be obtained by sampling the initial state $\ket{\phi}$ in some
Pauli-rotated basis.  In this case, the shot noise error incurred in the
computation $\mel{\phi}{P_k}{\phi}$ can be bounded by using a
result from \cite{huang2019nearterm}, which we reproduce here for convenience:
\begin{theorem}{\normalfont(Adapted from \cite{huang2019nearterm}.)}
Let $\epsilon > 0$ and $P_k$ be a tensored-Pauli operator over $n$
qubits. Let multiple copies of an arbitrary $n$-qubit quantum state
$\ket{\phi}$ be given. The expectation value $\mel{\phi}{P_k}{\phi}$ can be
determined to additive accuracy $\epsilon$ with failure probability at most
$\delta$ using $\mathcal{O}(\frac{1}{\epsilon^2}\log{(\frac{1}{\delta})})$
copies of $\ket{\phi}$.
\end{theorem}
The final measurement error can be somewhat mitigated by calibrating the POVM
matrix, which rotates from an ideal set of counts to one affected by
measurment noise.  This can be done using the \emph{"CompleteMeasFitter"}
function in the IBMQ circuit library.

To summarise thus far, the three sources of error mentioned above are
ultimately captured as errors in the matrix elements of the $D$ and $E$
matrices. This means that at the end of the day, we just need to look at the
$D$ and $E$ matrices. Then, errors in the $E_{ij}$ and $D_{ij}$ matrix
elements are propagated in our algorithm firstly through \eqref{eqn:
generalised eigenvalue problem}, which we use to determine the generalised
eigenvalues $\lambda_i$ and the corresponding eigenvectors $\vec{v}_i$.
I.e, these errors in the $D_{ij}$ and $E_{ij}$ matrix elements lead to errors
in the generalised eigenvalues $\lambda_i$ and the generalised eigenvectors
$\vec{v}_i$.  Subsequently, errors in $\lambda_i$ and $\vec{v}_i$
would lead to errors in $\vec{\alpha}(T)$ through \eqref{eqn: fast forwarding
time evo}.  We will now make these statements more precise.

Let $E,D$ be the ideal, noiseless $E$ and $D$ matrices and let $\tilde{E}$, $\tilde{D}$ be the matrices whose
matrix elements are computed on the quantum computer.  Let the corresponding
generalised eigenvalue problems be $D\vec{v} = \lambda E \vec{v}$
and $\tilde{D}\tilde{\vec{v}} = \tilde{\lambda} \tilde{E}
\tilde{\vec{v}}$ respectively. If we denote the noiseless CQFF result
as
\begin{subequations}
  \begin{align}
    \ket{\psi(\vec{\alpha}(T))} &= \sum_i \alpha_i(T) \ket{\chi_i}\\
    \vec{\alpha}(T) &= \sum_j e^{-i\lambda_j T}\vec{v}_j \vec{v}_j^\dagger E
    \vec{\alpha}(0)
  \end{align}
\end{subequations}
and the noisy CQFF result as 
\begin{subequations}
  \begin{align}
    \ket{\psi(\tilde{\vec{\alpha}}(T))} &= \sum_i \tilde{\alpha_i}(T)
    \ket{\chi_i}\\
    \tilde{\vec{\alpha}}(T) &= \sum_j e^{-i\tilde{\lambda}_j
    T}\tilde{\vec{v}}_j \tilde{\vec{v}}_j^\dagger \tilde{E}
    \vec{\alpha}(0)
  \end{align}
\end{subequations}
then we have
\begin{align}
  \braket{\psi(\vec{\alpha}(T))}{\psi(\tilde{\vec{\alpha}}(T))} 
  &= \vec{\alpha}(T)^\dagger E \tilde{\vec{\alpha}}(T) \nonumber \\
  &= \vec{\alpha}(0)^\dagger K \vec{\alpha}(0)
\end{align}
where 
\begin{equation}
  K = \sum_{j j^\prime} e^{-i(\tilde{\lambda}_{j^\prime} -\lambda_j)T} E
  \vec{v}_j \vec{v}_j^\dagger E
  \tilde{\vec{v}}_{j^\prime}\tilde{\vec{v}}_{j^\prime}^\dagger\tilde{E}.
\end{equation}
Hence, we see that the error in $\lambda_i$ hence leads to an error that is
periodic in time, whereas the error in $\vec{v}_i$ lead to an error that
is constant in the simulation time $T$ except when at $T=0$, where the error
vanishes since $\sum_i \tilde{\vec{v}}_i \tilde{\vec{v}}_i^\dagger
E = I$.

Now, the goal is to find a bound for the difference in eigenvalues
$|\lambda_i - \tilde{\lambda}_i|$ and also to find a way to quantify the
difference in the corresponding eigenspaces with the eigenvectors
$\vec{v}_j$ and $\tilde{\vec{v}}_j$.  To that end, we can apply
results from the well-studied field of matrix perturbation theory
\cite{bhatia2007perturbation,stewart1990matrix}.
Namely, if certain conditions hold for the $D$ and $E$ matrices, then there
are certain bounds that can be established.  We shall just give a few
examples below to demonstrate what we mean.
\begin{theorem}{\normalfont (Adapted from \cite{STEWART197969}.)}
  \label{theorem: stewart theorem 1}
  Suppose the $E$ and $\tilde{E}$ matrices are positive definite, which
  allows them to admit the Cholesky decomposition $E = R^\dag R$ and
  $\tilde{E} = \tilde{R}^\dag \tilde{R}$.  Define $A = (R^\dagger)^{-1} D
  R^{-1}$ and $\tilde{A} = (\tilde{R}^\dagger)^{-1} \tilde{D}
  \tilde{R}^{-1}$, which allows us to write the generalised eigenvalue
  problems $D\vec{v} = \lambda E \vec{v}$ and $\tilde{D}\tilde{\vec{v}} =
  \tilde{\lambda} \tilde{E} \tilde{\vec{v}}$ as $A\vec{\beta}_i = \lambda_i
  \vec{\beta_i}$ and $\tilde{A}\tilde{\vec{\beta}}_i = \tilde{\lambda}_i
  \tilde{\vec{\beta}}_i$.  Here, $\vec{\beta} = R\vec{v}$ and
  $\tilde{\vec{\beta}}=\tilde{R}\tilde{\vec{v}}$.
  Without loss of generality, we order the eigenvalues such that
  $\lambda_i \leq \lambda_j$, $\tilde{\lambda}_i \leq \tilde{\lambda}_j$ for
  $i<j$. Then, we have:
  \begin{equation}
    |\lambda_i - \tilde{\lambda}_i| \leq \norm*{A - \tilde{A}}
  \end{equation}
  where $\norm*{A - \tilde{A}}$ denotes the spectral norm.
\end{theorem}
We see that the trick here is to convert the two generalised eigenvalue
problems into normal hermitian eigenvalue problems, before using the
well-studied bound for normal hermitian eigenvalue problems.  The condition
for $E$ and $\tilde{E}$ to be positive definite might seem very strict at
first glance, but there are two things that must be said here.  Firstly, if
the set $\mathbb{C}\mathbb{S}_K$ is linearly independent, then the $E$ matrix
would be positive definite.  If the $\tilde{E}$ matrix is small enough, we
can directly check if it is positive definite too before applying this
theorem.  Secondly, as was mentioned in Appendix~\ref{Appendix: derivation of
diagonal representation}, we can ignore the action of $E$ and $\tilde{E}$ on
their corresponding nullspaces when using the CQFF algorithm.  This means
that it is possible to restrict ourselves to work in the column spaces of the
$E$ and $\tilde{E}$ matrices respectively, and in that restricted space, $E$
and $\tilde{E}$ are positive definite.  If we want to relax the above
condition of $E$ and $\tilde{E}$ being positive definite, we can use an
alternative bound for the eigenvalues, known as the Bauer-Fike
theorem\cite{bauer1960norms}, which we adapt below for the convenience of the
reader.
\begin{theorem}{\normalfont (Adapted from \cite{bauer1960norms}.)}
  \label{theorem: bauer-fike theorem}
  Let $E$ be a Hermitian matrix.  For each eigenvalue $\tilde{\lambda}_j$ of
  $\tilde{E}$, there is an eigenvalue $\lambda$ of $E$ such that
  \begin{equation}
    |\tilde{\lambda}_j - \lambda| < \norm*{\tilde{E}-E}.
  \end{equation}
  where here the norm $\norm*{E-\tilde{E}}$ is the spectral norm.
\end{theorem}
Apart from theorem~\ref{theorem: stewart theorem 1} and theorem~\ref{theorem:
bauer-fike theorem}, there are other bounds for the eigenvalues that can be
found in \cite{STEWART197969}.

To bound the difference in the various eigenspaces, for the case where both
$\tilde{E}$ and $E$ are positive definite, we can use the same transformation
as in theorem~\ref{theorem: stewart theorem 1} to convert the two generalised
eigenvalue problems $D\vec{v} = \lambda E\vec{v}$ and
$\tilde{D}\tilde{\vec{v}} =
\tilde{\lambda}\tilde{E}\tilde{\vec{v}}$ into two regular Hermitian
eigenvalue problems before using the well known Davis-Kahan $\sin(\Theta)$
theorem \cite{davisKahanTheorem} to provide bounds on the eigenspaces
corresponding to different eigenspaces of the regular Hermitian eigenvalue
problems. For completeness, we will also adapt the Davis-Kahan $\sin(\Theta)$
theorem here. 
\begin{theorem}{\normalfont (Adapted from \cite{davisKahanTheorem}.)}
  Suppose the $E$ and $\tilde{E}$ matrices are positive definite, which
  allows them to admit the Cholesky decomposition $E = R^\dag R$ and
  $\tilde{E} = \tilde{R}^\dag \tilde{R}$.  Define $A = (R^\dagger)^{-1} D
  R^{-1}$ and $\tilde{A} = (\tilde{R}^\dagger)^{-1} \tilde{D}
  \tilde{R}^{-1}$, which allows us to write the generalised eigenvalue
  problems $D\vec{v} = \lambda E \vec{v}$ and $\tilde{D}\tilde{\vec{v}} =
  \tilde{\lambda} \tilde{E} \tilde{\vec{v}}$ as $A\vec{\beta}_i = \lambda_i
  \vec{\beta_i}$ and $\tilde{A}\tilde{\vec{\beta}}_i = \tilde{\lambda}_i
  \tilde{\vec{\beta}}_i$.  Here, $\vec{\beta} = R\vec{v}$ and
  $\tilde{\vec{\beta}}=\tilde{R}\tilde{\vec{v}}$.
  Without loss of generality, let $A$ and $\tilde{A}$ be $n$ by $n$ matrices.
  Also, treat the vectors $\vec{\beta}$ and $\tilde{\vec{\beta}}$ as column
  matrices.
  Define
  \begin{align}
    V_0 &= \begin{pmatrix} \vec{\beta}_1 \dots \vec{\beta}_l \end{pmatrix}\\
    A_0 &= \text{diag}\begin{pmatrix}\lambda_1 \dots \lambda_l \end{pmatrix}\\
    V_1 &= \begin{pmatrix} \vec{\beta}_{l+1} \dots \vec{\beta}_n \end{pmatrix}\\
    A_1 &= \text{diag}\begin{pmatrix}\lambda_{l+1} \dots \lambda_n \end{pmatrix}
  \end{align}
  such that we have $A = V_0 A_0 V_0^\dagger + V_1 A_1 V_1^\dagger$.  Also
  define the terms $\tilde{V}_0,\tilde{A}_0,\tilde{V}_1,\tilde{A}_1$
  analogously such that we have $\tilde{A} = \tilde{V}_0 \tilde{A}_0
  \tilde{V}_0^\dagger + \tilde{V}_1 \tilde{A}_1 \tilde{V}_1^\dagger$.  Then,
  if the eigenvalues of $A_0$ are contained in an interval $(a,b)$ and the
  eigenvalues of $\tilde{A}_1$ are excluded from the interval $(a-\delta,
  b+\delta)$ for some $\delta > 0$, then
  \begin{equation}
    \norm*{V_0^\dagger \tilde{V}_1} \leq \frac{\norm*{V_0^\dagger(\tilde{A}-A)\tilde{V}_1}}{\delta}
  \end{equation}
  for any unitarily invariant norm $\norm*{\cdot}$.
\end{theorem}
A corresponding bound for the case where either $E$ is not positive definite
or $\tilde{E}$ is not positive definite can be found in \cite{STEWART197969}.

\section{Tweaking CQFF to make it more similar to VFF}
\label{Appendix: tweaking CQFF to look like VFF}
Define $U(\Delta t) = e^{-i H \Delta t}$. If $\Delta t$ is small, we have
\begin{align}
  \label{eqn: U(Delta t)}
  U(\Delta t) \approx \mathbbm{1} - i H \Delta t
\end{align}
and hence 
\begin{equation}
  \label{eqn: U(N Delta t)}
  U(N\Delta t) = U(\Delta t)^N \approx \left(\mathbbm{1} - i H \Delta t \right)^N \,.
\end{equation}
Note that $U(N\Delta t)$ commutes with $H$, and hence starting from
\eqref{eqn: H w.r.t CSK}, we define $[U(N\Delta
t)]_{\mathbb{C}\mathbb{S}_K}$ through
\begin{equation}
   [U(N\Delta t)]_{\mathbb{C}\mathbb{S}_K} = \sum_j \left(1 -
   i\lambda_j\Delta t\right)^N \vec{v}_j \vec{v}_j^\dag E
\end{equation}
where $\lambda_j$, $\vec{v}_j$ and $E$ have the same meaning as they do in
\eqref{eqn: H w.r.t CSK}. As mentioned in the main text, this method is
generally not preferable as compared with the original CQFF algorithm, as the
approximations in \eqref{eqn: U(Delta t)}~and~\eqref{eqn: U(N Delta t)}
introduce errors that can be avoided. Furthermore, time evolution with
\eqref{eqn: U(N Delta t)} is not norm-preserving, $U(N \Delta t)$ as defined
above is not unitary, but only approximately unitary. 

\section{Initial state preparation}
\label{Appendix: initial state preparation}
For the runs on IBM's quantum computer, we prepared a state $\ket{\psi}$ with
randomised parameters which we use for the CQFF algorithm. Using the
\emph{efficientSU2} function in IBMQ's Qiskit's circuit library, this initial
state was produced by 5 layers of gates on $\ket{00...0}$, where each layer
comprises of $SU(2)$ operations with randomised rotation angles on all the
qubits followed by CNOT gates to entangle all the qubits. \\

\section{Size of subspaces used in the CQFF algorithm for the different
Hamiltonians considered}
\label{Appendix: subspace sizes}
In Table~\ref{table: subspace size}, we tabulate the size of the subspaces
used in the CQFF algorithm for the different Hamiltonians considered.
\begin{table}[h]
\centering
\begin{tabular}{|l|l|l|l|}
\hline
                 & $K=1$ & $K=2$ & $K=3$ \\ \hline
$H_1$ (2 qubits) & 4     &       &       \\ \hline
$H_1$ (3 qubits) & 7     & 16    &       \\ \hline
$H_2$ (4 qubits) & 3     & 4     &       \\ \hline
$H_2$ (5 qubits) & 4     & 7     & 8     \\ \hline
$H_3$ (2 qubits) & 3     &       &       \\ \hline
\end{tabular}
\caption{Comparison of the sizes of $\mathbb{C}\mathbb{S}_K$ for the
different Hamiltonians considered in this paper.}
\label{table: subspace size}
\end{table}

\section{Supplementary plots}
\label{Appendix: Supplementary plots}
\subsection{Expectation value plot for $H_1$ with two qubits}
\label{Appendix: expectation value plot}
\begin{figure}[h]
  \centering
  \includegraphics[width=0.27\textwidth, trim={0cm 0cm 0cm 0cm},clip]{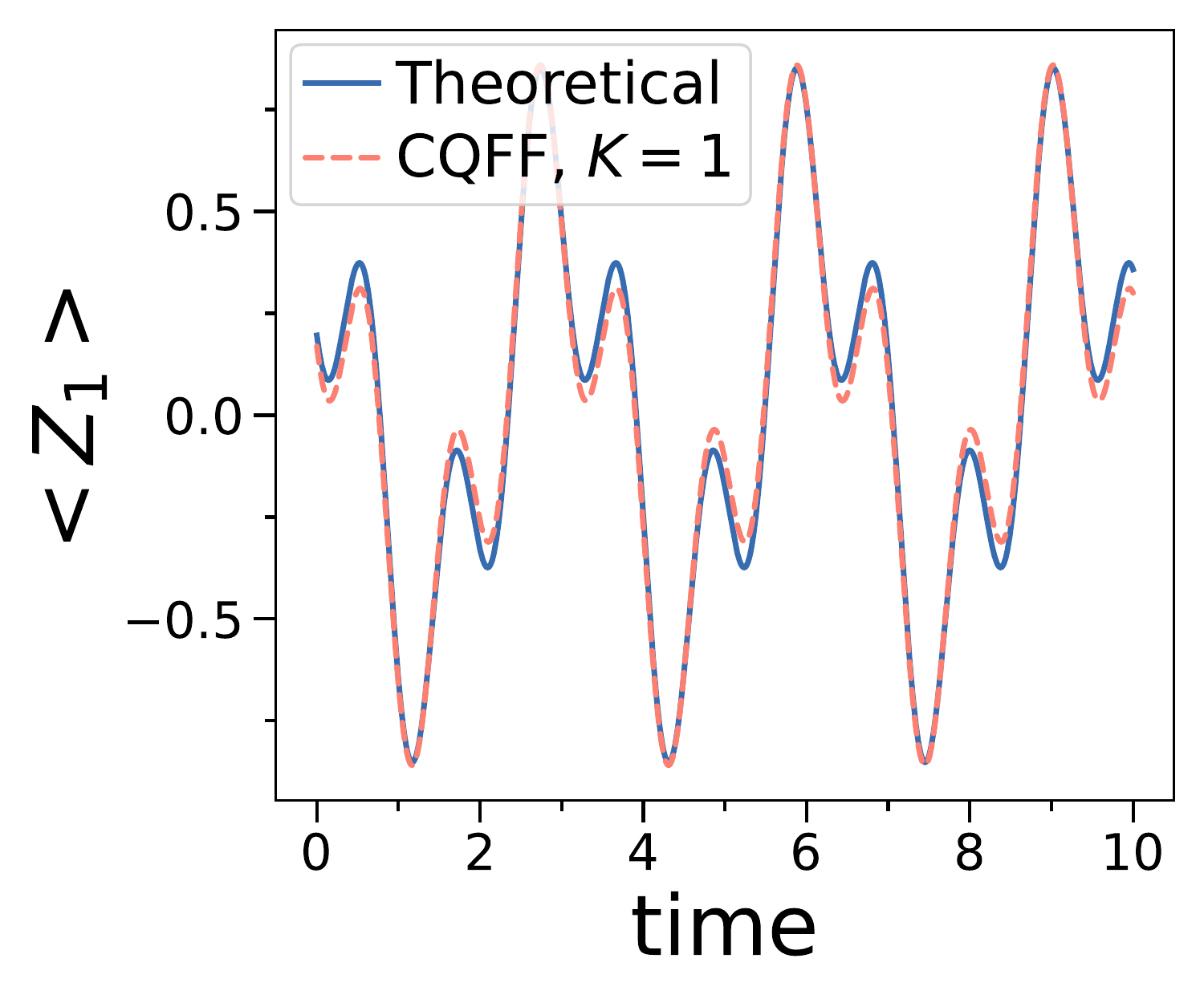}
  \caption[]{Time evolution of CQFF on a 2 qubit state with
  Hamiltonian $H_1$, simulated on the IBM quantum processor
  \emph{ibmq\_lagos} with $8192$ shots. The expectation value $\langle Z_1
  \rangle$ is shown here. Here, as compared to \figref{fig: 2 qubit XYZ
  results}, terms like $\mel{\chi_i}{O}{\chi_j}$ are computed directly on the
  quantum computer. }
  \label{fig: 2 qubit XYZ results real qm com}
\end{figure}

As mentioned in the main text we compute an observable $O$ as following
\begin{align}
  \langle O(t) \rangle_{CQFF} 
  &= \mel{\psi(t)_{CQFF}}{O}{\psi(t)_{CQFF}} \nonumber \\
  &= \sum_i^L \sum_j^L \alpha_i(t)^*
  \mel{\chi_i}{O}{\chi_j}\alpha_j(t)
\end{align}
The coefficients $\vec{\alpha}(T)$ as defined in
\eqref{eqn: cqff state} are computed using a quantum computer by measuring the $D$ and $E$ matrices. These crucial terms describe the evolution of the quantum state in time. 
To calculate explicit expectation values $\langle O(t) \rangle_{CQFF}$, we additionally need terms like $\mel{\chi_i}{O}{\chi_j}$.  In the main text, these terms  $\mel{\chi_i}{O}{\chi_j}$ are computed
using a classical computer.  Here for completeness, we plot the $\langle Z_1(t)
\rangle$ for the $2$ qubit case of $H_1$ where the terms
$\mel{\chi_i}{Z_1}{\chi_j}$ are computed directly on the quantum computer by
sampling the state $\ket{\phi}$ in the corresponding Pauli-rotated basis.
Our results are shown in \figref{fig: 2 qubit XYZ results real qm com}.  The
computation of $\mel{\chi_i}{O}{\chi_j}$ on the quantum computer introduces a
small error in the time evolution due to the noise of the quantum computer.

\subsection{Fidelities at long time scales}
\label{Appendix: long time fidelity plots}
\begin{figure}[H]
  \centering
  \subfigimg[width=0.23\textwidth]{a}{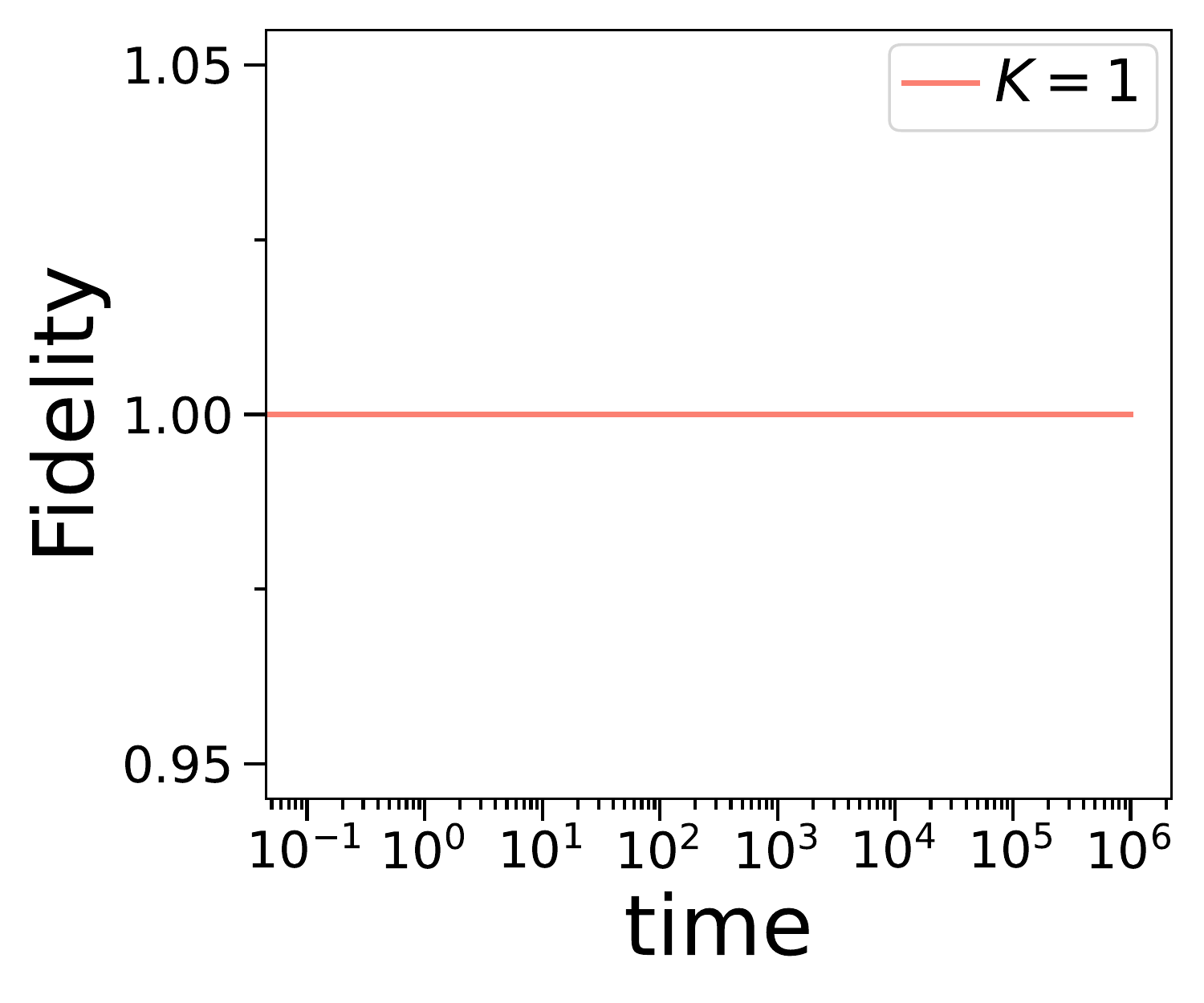}\hfill
  \subfigimg[width=0.23\textwidth]{b}{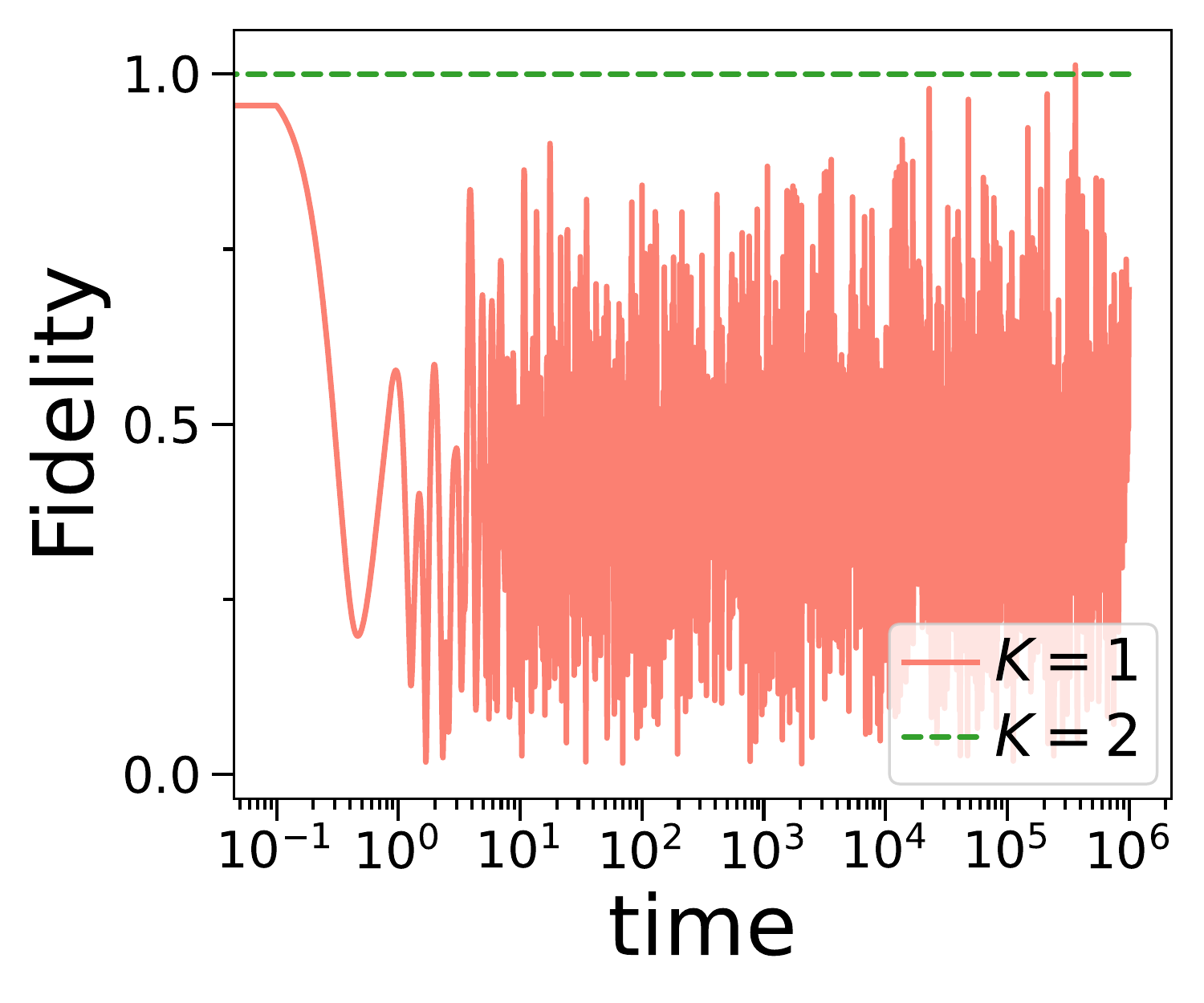} \\
  \subfigimg[width=0.23\textwidth]{c}{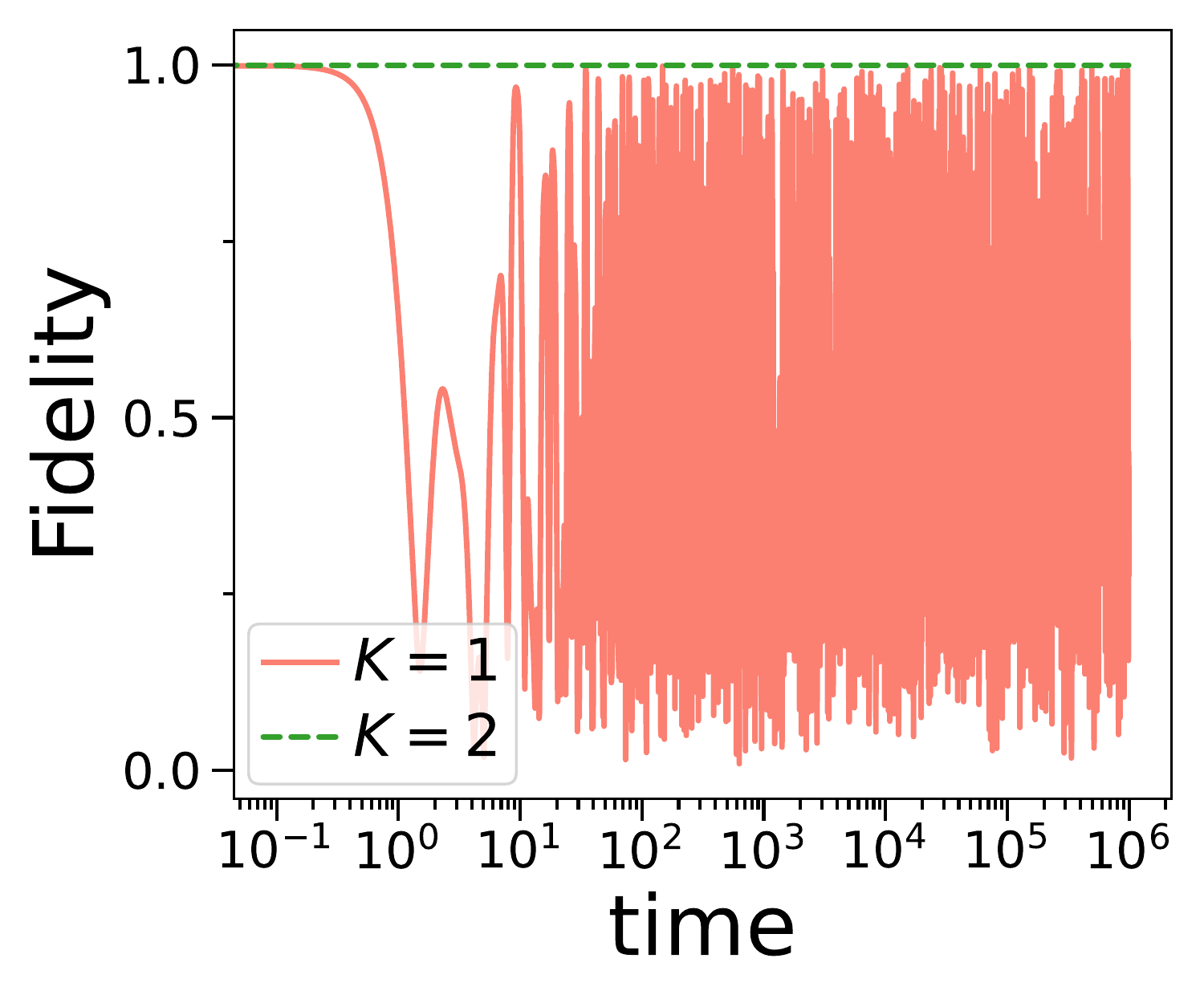} \hfill
  \subfigimg[width=0.23\textwidth]{d}{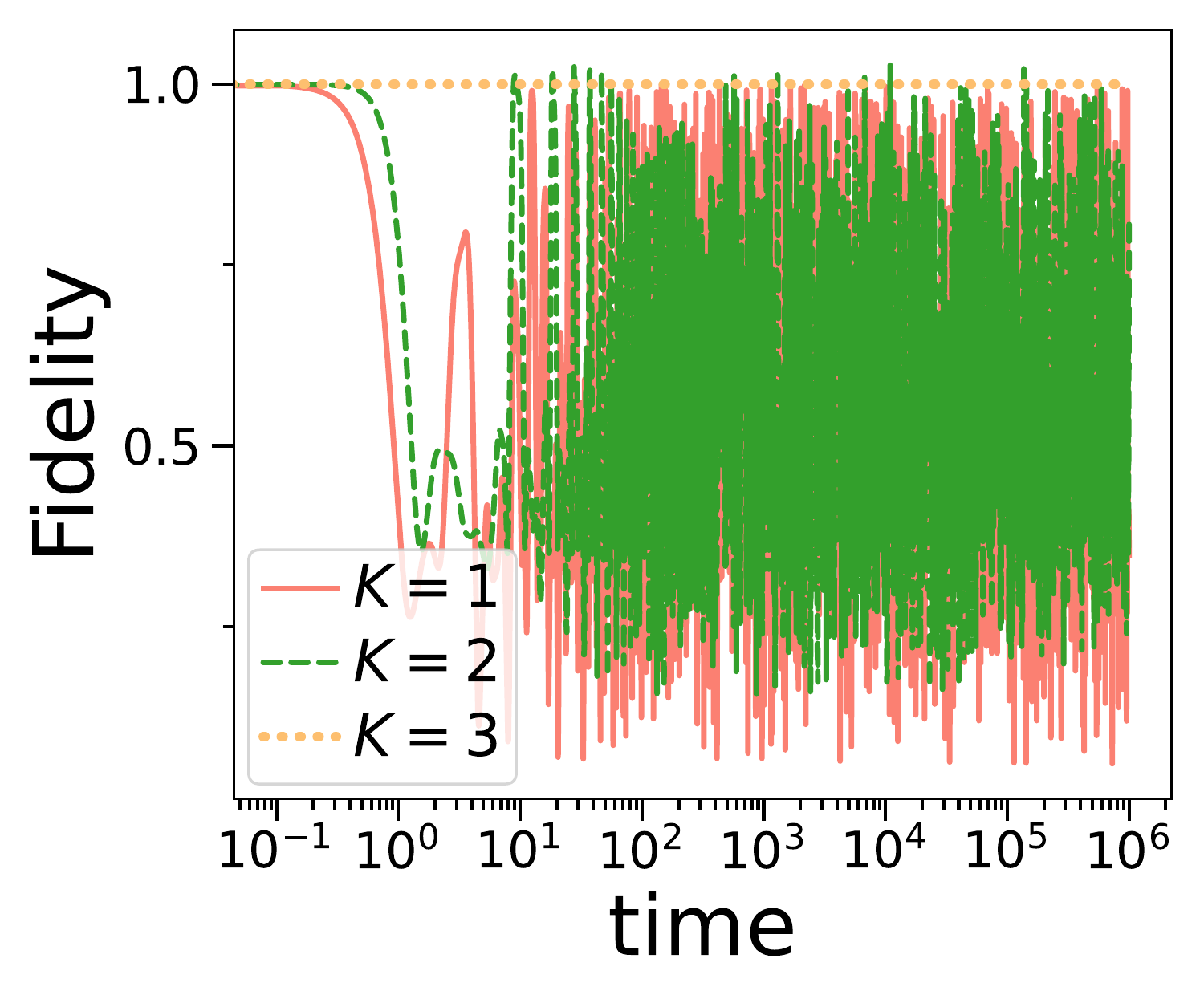}
  \caption[]{The long term behaviour of the fidelities for the Hamiltonians
  considered in the main text. As can be seen, for sufficiently high $K$, the
  fidelity essentially remains at $1$. For lower values of $K$, the
  fidelities remain oscillatory, as reflected in the log plots below. \\
  \textbf{a)} Long term behavior of the
  fidelity for the Hamiltonian in \figref{fig: 2 qubit XYZ
  results}.
  \textbf{b)} Long term behavior of the fidelity for the
  Hamiltonian in \figref{fig: 3 qubit XYZ results}. 
  \textbf{c)} Long term
  behavior of the fidelity for the Hamiltonian in \figref{fig: 4 qubit
  three-body results}.
  \textbf{d)} Long term behavior of the fidelity for
  the Hamiltonian in \figref{fig: 5 qubit three-body results}.}
  \label{fig: combined long term fidelity plots}
\end{figure}

In \figref{fig: combined long term fidelity plots}, we show the dynamics of
the fidelities for long time scales for the Hamiltonians considered in the
main text.

\section*{Acknowledgements}
We are grateful to the National Research Foundation and the Ministry of
Education, Singapore for financial support. The authors acknowledge the use
of the IBM Quantum Experience devices for this work. This work is supported
by a Samsung GRP project and the UK Hub in Quantum Computing and Simulation,
part of the UK National Quantum Technologies Programme with funding from UKRI
EPSRC grant EP/T001062/1.

\section*{Data availability}
The authors declare that the main data supporting the findings of this study are available within the article. Extra data sets are available upon request.

\bibliographystyle{apsrev4-1}
\bibliography{CQFF_paper}
\end{document}